\documentclass[10pt]{article}
\usepackage{makeidx}

\def \Oeuvres{O$\!$euvres}
\def\LHS{lhs~}
\def \D {\hbox{d}}
\def \Log {\mathop{\rm Log}\nolimits}
\def \sech {\mathop{\rm sech}\nolimits}

\def \Im  {\mathop{\rm Im}\nolimits}
\def \arg {\mathop{\rm arg}\nolimits}
\def \mod#1{\vert #1 \vert}
\def \jmax{J}
\def\JPSJ{J.~Phys.~Soc.~Japan}
\def\PTP{Prog.~Theor.~Phys.~}
\def\GLA{A}

\def\pcarre{{\vert p \vert}^2}

\begin{document}

\title{Solitary waves of nonlinear nonintegrable equations
\footnote{
\textit{Dissipative solitons},
ed.~N.~Akhmediev,
Lecture notes in physics
(Springer, Berlin, 2004). S2003/099
}}

\author
{
Robert CONTE~$^\dag$ 
and 
Micheline MUSETTE~$^\ddag$
\\
$^\dag$ Service de physique de l'\'etat condens\'e
%(Unit\'e de recherche associ\'ee au CNRS no.~2464) 
(URA no.~2464) 
\\~~CEA--Saclay, F--91191 Gif-sur-Yvette Cedex, France 
\\~~E-mail: Conte@drecam.saclay.cea.fr\\[10pt]
$^\ddag$ Dienst Theoretische Natuurkunde, Vrije Universiteit Brussel
\\~~Pleinlaan 2, B--1050 Brussels, Belgium
\\~~E-mail: MMusette@vub.ac.be
}

\maketitle

\hfill % \today

{\vglue -10.0 truemm}
{\vskip -10.0 truemm}

\begin{abstract}
Our goal is to find closed form analytic expressions
for the solitary waves of nonlinear
nonintegrable partial differential equations.
The suitable methods, which can only be nonperturbative,
are classified in two classes.

In the first class,
which includes the well known so-called truncation methods,
one \textit{a priori} assumes a given class of expressions
(polynomials, etc) for the unknown solution;
the involved work can easily be done by hand
but all solutions outside the given class are surely missed.

In the second class,
instead of searching an expression for the solution,
one builds an intermediate, equivalent information,
namely the \textit{first order}
autonomous ODE satisfied by the solitary wave;
in principle, no solution can be missed,
but the involved work requires computer algebra.

We present the application to
the cubic and quintic complex one-dimensional Ginzburg-Landau equations,
and to the Kuramoto-Sivashinsky equation.

\end{abstract}

\noindent \textit{Keywords}:
solitary waves,
complex one-dimensional Ginzburg-Landau equation,
Kuramoto-Sivashinsky equation,
complex Swift-Hohenberg equation,
Briot and Bouquet equations,
elliptic function,
genus,
Painlev\'e property,
meromorphy,
truncation.

\noindent \textit{PACS 1995}~:
% (cosmology, general relativity, nonlinear ODEs, quantum gravity)
 02.30.-f   % Math. methods in phys; Function theory, analysis
%02.60.-x   % Math. methods in phys; Numerical approx. and analysis
%02.70.-c   % Math. methods in phys; Computational techniques
 05.45.+b   % Stat phys and thermo; Theory and models of chaotic systems
 42.65.-k   % Fundam. areas of phenomenology; Nonlinear optics
 47.27.-i   % Fundam. areas of phenomenology; Turbulent flows, convection, ...
%74.20.-z   % Condensed matter:...; Superconductivity; Theory
%82.40.-g   % Cross-discipl...; Chemical kinetics and reaction (.Py = flames)

%baselineskip= 8truept
%baselineskip=10truept % 64 pages
\baselineskip=12truept % 68 pages

% FOR DOUBLE-SPACING, REMOVE THE TWO FOLLOWING %
%\baselineskip=24truept
%\renewcommand{\baselinestretch}{2.0}

\tableofcontents

\vfill \eject

% ============================================================================
\section{Introduction}

Many nonlinear partial differential equations (PDEs) encountered in physics
are autonomous, i.e. do not depend explicitly on the independent variables
$x$ (space) and $t$ (time).
In such a case,
they admit a reduction, called traveling wave reduction,
to an autonomous nonlinear ordinary differential equation (ODE),
defined in the simplest case by $u(x,t)=U(\xi), \xi=x-ct$,
with $c$ a constant speed.
A \textit{solitary wave} is then defined as any solution
to this nonlinear autonomous ODE.
Physically relevant solitary waves must satisfy some decaying condition
when $\xi$ goes to $\pm \infty$.

The distinctive feature of this chapter is to explain the methods to find
closed form expressions to these solitary waves
when the PDE and the reduced ODE are algebraic and nonintegrable.
These solitary waves may have the topology of
a front (for instance $\tanh$),
a pulse (for instance $\sech$),
a source, a sink, etc,
but we will not discard an apparently physically uninteresting solution,
because it might appear interesting to another field.

Why ``algebraic''?
This only excludes equations impossible to convert to an algebraic form.
For instance, the sine-Gordon equation $u_{xt}-\sin u=0$ is not excluded
because it is algebraic in $e^{i u}$.

Why ``nonintegrable''?
Because the integrable ones (nonlinear Schr\"odinger (NLS),
coupled NLS in the Manakov case, etc) are ``easy'' to solve using powerful
tools like the inverse spectral transform (IST) \cite{AblowitzClarkson}.
The difficulty with the nonintegrable equations
is the absence of a general method to achieve the goal.

Why ``autonomous''?
Irrelevant for the truncation methods
(section \ref{sectionFirstClass}),
this restriction is essential
for the mathematical method of section \ref{sectionSecondClass}.
Physically, this is not an important restriction,
since many interesting PDEs are autonomous,
see the examples below.

Why ``closed form expressions''?
Because a solution represented by a series can be misleading
(\textit{illusoire}, used to say Painlev\'e).
Consider for instance a chaotic deterministic dynamical system
for which no analytic solution exists.
Around a regular point,
it admits a solution represented by a Taylor series,
but one can conclude nothing before some analytic continuation has been
performed.
On the contrary, the Laurent series around a movable singularity
(i.e.~one whose location depends on the initial conditions)
provides some constructive information 
(see section \ref{sectionSingularityAnalysis})
about the (global) integrability of the equation.

The methods described here are all based on the \textit{a priori}
singularities \cite{Cargese1996Musette} 
of the solutions of the given ODE.
In particular, we do not consider the group theoretical methods
\cite{Cargese1996ClarksonWinternitz}.

These methods can be applied mainly to dissipative equations of
importance in physics, nonlinear optics, mechanics, etc.
Our specific examples are the following.

\begin{enumerate}
%                           ******************************
\item
\index{complex Ginzburg--Landau!(CGL3) equation}
The one-dimensional cubic complex Ginzburg-Landau equation (CGL3)
\begin{eqnarray}
{\hskip -10.0 truemm}
& &
i A_t + p A_{xx} + q \mod{A}^2 A - i \gamma A =0,\
p q \gamma \not=0,\
\Im(p/q)\not=0,\
(A,p,q) \in {\mathcal C},\
\gamma  \in {\mathcal R},
\label{eqCGL3}
\end{eqnarray}
(and its complex conjugate, i.e.~a total differential order four),
in which $p,q,\gamma$ are constants,
a generic equation which describes many physical phenomena,
such as the propagation of a signal in an optical fiber
\cite{AgrawalBook},
spatiotemporal intermittency in spatially extended dissipative systems
\cite{MannevilleBook,vHSvS,vS2003}.
We will restrict ourselves to the CGL3 case properly said
$\Im(p/q)\not=0$.

For analytic results on two coupled CGL3 equations,
see \cite{CM2000b}.

%                           ******************************
\item
The Kuramoto and Sivashinsky (KS) equation,
\index{Kuramoto--Sivashinsky (KS)!equation}
\begin{eqnarray}
& &
\varphi_t + \nu \varphi_{xxxx} + b \varphi_{xxx} + \mu \varphi_{xx}
 + \varphi \varphi_x = 0,\quad
    \varphi \in \mathcal{C},\,
    (\nu,b,\mu) \in \mathcal{R},\
\nu \not=0.
\label{eqKS}
\end{eqnarray}
in which $\nu,b,\mu$ are constants.
This PDE is obeyed by the variable $\varphi=\arg A$ of the above field $A$
of CGL3 under some limit \cite{PM1979,Lega2001},
hence its name of phase turbulence equation.

%                           ******************************
\item
The one-dimensional quintic complex Ginzburg-Landau equation (CGL5),
\begin{eqnarray}
{\hskip -10.0 truemm}
& &
i A_t +p A_{xx} +q \mod{A}^2 A +r \mod{A}^4 A -i \gamma A =0,\
p r \not=0,\
\Im(p/r)\not=0,\
(A,p,q,r) \in {\mathcal C},\
\gamma  \in {\mathcal R}.
\label{eqCGL5}
\end{eqnarray}

%                           ******************************
\item
The Swift-Hohenberg equation \cite{SH1977,LMN}
\index{Swift-Hohenberg equation}
\begin{eqnarray}
{\hskip -10.0 truemm}
& &
i A_t +b A_{xxxx} +p A_{xx} +q \mod{A}^2 A +r \mod{A}^4 A -i \gamma A =0,\
b r \not=0,\
(A,b, p,q,r,) \in {\mathcal C},\
\gamma  \in {\mathcal R},
\label{eqSH}
\end{eqnarray}
in which $b,p,q,r,\gamma$ are constants.

%                           ******************************
\end{enumerate}

For the CGL3, KS, and Swift-Hohenberg equations
(with one exception, KS with $b^2=16 \mu \nu$),
all the solitary wave solutions
$\mod{A}^2=f(\xi), % \mod{B}^2=g(\xi),
\varphi=\Phi(\xi), \xi=x-ct$,
which are known hitherto are just polynomials in $\tanh k \xi$.

So, two natural questions arise:

\begin{enumerate}
\item
Is it possible or impossible that other solitary waves exist?

\item
If such other solutions may exist,
can one find not just a few more but all of them?

\end{enumerate}

Let us from now on denote the reduced ODE as
\begin{eqnarray}
& &
E(u^{(N)},...,u',u)=0,\
'=\frac{\D}{\D \xi},\
\xi=x-ct.
\label{eqODEReduced}
\end{eqnarray}
and let us assume that it is also nonintegrable.

The chapter is organized as follows.

In section \ref{sectionKnownSolutions},
one recalls the analytic expressions of the known
solitary waves of the examples.
This list,
to be retrieved or augmented by the singularity based methods,
will allow us to rate the efficiency of the various methods.

In section \ref{sectionCount},
one investigates the amount of integrability of the equation,
by applying the so-called Painlev\'e test.
More specifically, one checks the existence of particular solutions
which admit a \textit{local} representation as a Laurent series.
This allows us to count the gap,
strictly positive because of the assumed nonintegrability,
between
the differential order of the ODE
and the maximal number of available integration constants.

In section \ref{sectionSelectionSinglevalued},
we discuss the choice of the suitable dependent variable
to be used in the subsequent sections.

In section \ref{sectionBothClasses},
one tries to obtain a \textit{global} representation
for the local information (Laurent series) previously found.
One introduces the distinction between two main classes of methods,
according to the following criteria:
i) computations easy enough to be carried out by hand,
ii) generality or particularity of the expected solution.

The section \ref{sectionFirstClass} is devoted to the first class of methods,
which are known as ``truncation methods''.
The input is a class of \textit{a priori} expressions for $u$
(usually polynomials),
in some intermediate variable $\chi$ which satisfies
a given first order ODE
(e.g. Riccati, Weierstrass, Jacobi).
Then,
by a direct computation,
easy to carry out by hand,
one checks whether there indeed exist solutions in the given class.
The solutions with a simple profile
(such as $\tanh$ for a front, $\sech$ for a pulse),
are easily found by this class of methods.

In the second class of methods \cite{MC2003},
presented in section \ref{sectionSecondClass},
rather than directly looking for the unknown solution
\begin{eqnarray}
& &
u = f(\xi-\xi_0),
\label{eqFormalSolution}
\end{eqnarray}
in which $\xi_0$ is an arbitrary complex constant,
one looks as an intermediate information for the first order nonlinear ODE
\begin{eqnarray}
& &
F(u,u') =0,
\label{eqOrder1Autonomous}
\end{eqnarray}
obtained by eliminating $\xi_0$ between
(\ref{eqFormalSolution}) and its derivative,
in which $F$ is as unknown as $f$.
Indeed, provided that $f$ is singlevalued,
by a classical theorem recalled in Appendix,
there is equivalence betwen the knowledge of the solution $f$
and that of the subequation $F$ which it satisfies.
\index{subequation}
The way to obtain the subequation $F$ is to require that it be
satisfied by the Laurent series obtained in a previous step.

The difference between the two classes of methods is the following.
The solutions found by the first class of methods can only be a subset
of those found by the second class.
However,
the computations involved can easily be performed by hand
for the first class,
while for the second class a computer algebra package
is highly recommended.

% ========================================================================
\section{The known solutions of the examples}
\label{sectionKnownSolutions}

None of the expressions listed below represents the largest analytic
solution which one could find,
and their distance to this largest, yet unknown, analytic solution
will be computed precisely in section \ref{sectionCount}.

% ========================================================================
\subsection{CGL3}
\label{sectionKnownSolutionsCGL3}

\index{complex Ginzburg--Landau!(CGL3) equation}

The traveling wave reduction of (\ref{eqCGL3})
\begin{eqnarray}
& &
A(x,t)=\sqrt{M(\xi)} e^{i(\displaystyle{-\omega t + \varphi(\xi)})},\
\xi=x-ct,\
(c,\omega,M,\varphi) \in {\mathcal R},
\label{eqCGL3red}
\\
& &
{\hskip -10.0 truemm}
 \frac{M''}{2 M} -\frac{{M'}^2}{4 M^2} + i \varphi''- {\varphi'}^2
        + i \varphi' \frac{M'}{M}
%- \frac{i}{p} \frac{c M'}{2 M}
- i \frac{c}{2p} \frac{M'}{M}
+ \frac{1}{p} \left(c \varphi' + \omega\right)
+ \frac{q}{p} M
- \frac{i \gamma}{p}
=0,
\label{eqCGL3ReducComplex}
\end{eqnarray}
introduces two additional real constants $(c,\omega)$
and it is convenient to define
the six real parameters $d_r,d_i,s_r,s_i,g_r,g_i$,
\begin{eqnarray}
& &
d_r + i d_i = \frac{q}{p},\
s_r - i s_i = \frac{1}{p},\
g_r + i g_i = \frac{\gamma + i \omega}{p} + \frac{1}{4} c^2 s_r^2.
\end{eqnarray}

In the CGL3 case properly said $d_i \not=0$ to which we restrict here,
only three % in the parameter space $g_r,g_i,c s_i,d_r,d_i$
% NDLR $c s_i=0$ always splits in two solutions
solutions are currently known.
Denoting $A_0^2$ and $\alpha$ two real constants defined by the
complex equation
\begin{eqnarray}
& &
(-1+i \alpha) (-2+i \alpha) p + A_0^2 q=0,
\label{eqCGL3LeadingOrderComplex}
\end{eqnarray}
these three solutions are the following.
\begin{enumerate}

\item
A heteroclinic source or propagating hole \cite{BN1985}
%[NDLR Ne rien mettre de r\'eel]
\begin{eqnarray}
& &
\left\lbrace
\matrix{
\displaystyle{
A=A_0 \left[\frac{k}{2} \tanh \frac{k}{2} \xi
 - \frac{ i q p_i}{2(1-i \alpha) p \pcarre d_i} c \right]
e^{\displaystyle
{i[\alpha \Log \cosh \frac{k}{2} \xi + \frac{q_i}{2 \pcarre d_i} c \xi
 - \omega t]}},
}
\hfill \cr
\displaystyle{
\frac{i \gamma- \omega}{p}=
\left(\frac{c}{2 p}\right)^2 - (2-3 i \alpha) \frac{k^2}{4},
}
\hfill \cr
}
\right.
\label{eqCGL3Hole}
\end{eqnarray}
in which the velocity $c$ is arbitrary.
Indeed, the real and imaginary parts of the last equation
define the value of $\omega$ and a linear relation between $c^2$ and $k^2$,
see \cite[Eq.~(79)]{CM1993}.

\item
A homoclinic pulse or solitary wave \cite{PS1977}
\begin{eqnarray}
& &
\left\lbrace
\matrix{
\displaystyle{
 A=A_0 (-i k \sech k x)
 e^{\displaystyle{i[\alpha \Log \cosh k x - \omega t ]}},
}
\hfill \cr
\displaystyle{
\frac{i \gamma- \omega}{p}=(1-i \alpha)^2 k^2,\
c=0.
}
\hfill \cr
}
\right.
\label{eqCGL3Pulse}
\end{eqnarray}

\item
A heteroclinic front or shock \cite{NB1984}
%NDLR Ne rien mettre de r\'eel]
\begin{eqnarray}
& &
\left\lbrace
\matrix{
\displaystyle{
A=A_0 \frac{k}{2} \left[\tanh \frac{k}{2} \xi + \varepsilon\right]
 e^{\displaystyle{i[\alpha \Log \cosh \frac{k}{2} \xi
      + \frac{3 p_r + \alpha p_i}{6 \pcarre} c \xi
      - \omega t]}},\
  \varepsilon^2=1,
}
\hfill \cr
\displaystyle{
\frac{i \gamma- \omega}{p}=
\left(\frac{c}{2 p}\right)^2 + \frac{k^2}{4},\
\frac{k}{2}=
\varepsilon \frac{p_i c}{6 \pcarre}.
}
\hfill \cr
}
\right.
\label{eqCGL3Front}
\end{eqnarray}

\end{enumerate}

None of these three solutions
requires any constraint on $p,q,\gamma$,
and they depend on an additional sign
resulting from the resolution of (\ref{eqCGL3LeadingOrderComplex}),
\begin{eqnarray}
& &
A_0^2 =\frac{3 (3 d_r + \varepsilon_1 \Delta)}{2 d_i^2},\
\alpha=\frac{   3 d_r + \varepsilon_1 \Delta} {2 d_i},\
\Delta=\sqrt{9 d_r^2 + 8 d_i^2},\
\varepsilon_1^2=1.
\label{eqCGL3LeadingOrderReal}
\end{eqnarray}
In all of them $M$ is a much simpler expression,
namely a second degree polynomial in
\begin{eqnarray}
\tau=(k/2) \tanh k \xi/2,\ k^2 \in \mathcal{R}.
\label{eqtau}
\end{eqnarray}
Therefore,
if one wants to extend the three above solutions,
it is advisable to eliminate $\varphi$ between the
system of two real equations equivalent to (\ref{eqCGL3ReducComplex}),
\begin{eqnarray}
& &
\left\lbrace
\begin{array}{ll}
\displaystyle{
\frac{M''}{2 M} -\frac{{M'}^2}{4 M^2} - {\varphi'}^2
- s_i \left(\frac{c M'}{2 M} + \gamma\right)
+ s_r \left(c \varphi' + \omega\right)
+ d_r M
=0,
}
\\
\displaystyle{
\varphi'' + \varphi' \frac{M'}{M}
- s_r \left(\frac{c M'}{2 M} + \gamma\right)
- s_i \left(c \varphi' + \omega\right)
+ d_i M
=0,}
\end{array}
\right.
\label{eqCGL3ReducRealSystem}
\end{eqnarray}
which results in
\begin{eqnarray}
& &
\varphi' = \frac{c s_r}{2} + \frac{G'-2 c s_i G}{2 M^2( g_r - d_i M)},\
\left(\varphi' - \frac{c s_r}{2}\right)^2=\frac{G}{M^2},
\label{eqCGL3Phiprime}
\\
& &
(G'-2 c s_i G)^2 - 4 G M^2  (d_i M - g_r)^2=0,\
\label{eqCGL3Order3}
\\
& &
G=\frac{1}{2} M M'' - \frac{1}{4} M'^2
  -\frac{c s_i}{2} M M' + d_r M^3 + g_i M^2,
\end{eqnarray}
and to concentrate on the single third order equation
(\ref{eqCGL3Order3}) for $M=\mod{A}^2$.

% ========================================================================
\subsection{KS}
\label{sectionKnownSolutionsKS}

\index{Kuramoto--Sivashinsky (KS)!equation}

The traveling wave reduction is defined as
\begin{eqnarray}
& &
\varphi(x,t)=c+u(\xi),\ \xi=x-ct,\
\left[\nu u''' + b u'' + \mu u' + \frac{u^2}{2}\right]' = 0,\
    (\nu,b,\mu) \in \mathcal{R},\
\nu \not=0,
\end{eqnarray}
which integrates once as
\begin{eqnarray}
& &
%E(u,\xi) \equiv
 \nu u''' + b u'' + \mu u' + \frac{u^2}{2} + A = 0,
%\nu u''' + b u'' + \mu u' + \frac{u^2}{2} - \frac{B^2}{2} = 0,
\label{eqKSODE}
\end{eqnarray}
in which $A$ in an integration constant.
It has a chaotic behavior \cite{MannevilleBook},
and it depends on two dimensionless parameters,
$b^2/(\mu \nu)$ and $\nu A / \mu^3$.

The known solutions are
one elliptic solution,
six trigonometric solutions,
and one rational solution.

The unique known elliptic solution exists for one constraint
between the parameters $\nu,b,\mu$ of the PDE
\cite{FournierSpiegelThual,Kud1990},
\begin{eqnarray}
& &
{\hskip -16.0 truemm}
b^2 - 16 \mu \nu=0 :\
u=-60\nu \wp' - 15b\wp - \frac{b \mu}{4 \nu},\
g_2=\frac{\mu^2}{12 \nu^2},\
g_3=\frac{13\mu^3+\nu A}{1080 \nu^3}.
\label{eqKSConstraint16}
\label{eqKSElliptic}
\end{eqnarray}
in which $\wp$ is the elliptic function of Weierstrass,
defined by the ODE
\begin{eqnarray}
& &
{\wp'}^2=4 \wp^3 - g_2 \wp - g_3.
\label{eqWeierstrass}
\end{eqnarray}

The six trigonometric solutions
\cite{KuramotoTsuzuki,KudryashovKSFourb},
all of them rational in $e^{k \xi}$,
exist at the price of one constraint between $\nu,b,\mu$
and another one on $A$,
\begin{eqnarray}
u & = &
 120 \nu \tau^3
 - 15 b \tau^2
 +\left(\frac{60}{19} \mu - 30 \nu k^2 - \frac{15 b^2}{4 \times 19 \nu}\right)
 \tau
 + \frac{5}{2} b k^2 - \frac{13 b^3}{32 \times 19 \nu^2}
  + \frac{7 \mu b}{4 \times 19 \nu},
\label{eqKSTrigo}
\\
\tau & = & \frac{k}{2} \tanh \frac{k}{2} (\xi-\xi_0),
\nonumber
\end{eqnarray}
the allowed values being listed in Table \ref{TableKS}.

\begin{table}[h] % [p]
\caption[garbage]{
The six known trigonometric solutions of KS, Eq.~(\ref{eqKSODE}).
They all have the form (\ref{eqKSTrigo}).
The last line is a degeneracy of the elliptic solution
(\ref{eqKSElliptic}).
}
\vspace{0.2truecm}
\begin{center}
\begin{tabular}{| c | c | c |}
%\hline % \hline % ********************************************************
\hline % \hline % ********************************************************
$b^2/(\mu\nu)$ & $\nu A/\mu^3$ & $\nu k^2/\mu$
\\ \hline \hline % ********************************************************
$0$ & $-4950/19^3,\ 450/19^3$ & $11/19,\ -1/19$
\\ \hline % \hline % ********************************************************
$144/47$ & $-1800/47^3$ & $1/47$
\\ \hline % \hline % ********************************************************
$256/73$ & $-4050/73^3$ & $1/73$
\\ \hline % \hline % ********************************************************
$16$ & $-18,\ -8$ & $1,\ -1$
\\ \hline % \hline % ********************************************************
\end{tabular}
\end{center}
\label{TableKS}
\end{table}

Finally, the unique known rational solution
\begin{eqnarray}
& &
b=0,\ \mu=0,\ A=0\:\ u = 120\nu (\xi-\xi_0)^{-3},
\end{eqnarray}
is a limit of all the above solutions.

A nice property common to all those solutions is to admit the representation
\begin{eqnarray}
u & = &
\mathcal{D} \Log \psi + \hbox{constant},\
\mathcal{D}=60 \nu \frac{\D^3 }{\D \xi^3} + 15 b \frac{\D^2 }{\D \xi^2}
           + \frac{15(16 \mu \nu - b^2)}{76 \nu} \frac{\D }{\D \xi},
\label{eqKSD}
%\\
%u_3 & = &
% \frac{5}{2} b k^2 - \frac{13 b^3}{32 \times 19 \nu^2}
%  + \frac{7 \mu b}{4 \times 19 \nu},\
\end{eqnarray}
in which $\psi$ is an \textit{entire} function
(i.e.~one without any singularity at a finite distance)
whose ODE is easy to build, respectively,
\begin{eqnarray}
& &
{(-\Log \psi)'''}^2 -4 {(-\Log \psi)''}^3 + g_2 (-\Log \psi)'' +g_3=0,
\\
& &
\psi'' - \frac{k^2}{4} \psi =0,
\\
& &
\psi''=0.
\end{eqnarray}
This linear operator $\mathcal{D}$,
which captures the singularity structure,
is called the
\textit{singular part operator}.
 \index{singular part operator}

% ========================================================================
\subsection{CGL5}
\label{sectionKnownSolutionsCGL5}

The traveling wave reduction is quite similar to that of CGL3,
so we do not repeat it.
Again, $A_0^2$ and $\alpha$ denote two real constants defined by the
complex equation
\begin{eqnarray}
& &
(-1/2+i \alpha) (-3/2+i \alpha) p + A_0^4 r=0,
\label{eqCGL5LeadingOrderComplex}
\end{eqnarray}
and the convenient constants are
\begin{eqnarray}
& &
e_r + i e_i = \frac{r}{p},\
s_r - i s_i = \frac{1}{p},\
g_r + i g_i = \frac{\gamma + i \omega}{p} + \frac{1}{4} c^2 s_r^2.
\end{eqnarray}

In the CGL5 case properly said $e_i \not=0$ to which we restrict here,
only two solutions are currently known.

\begin{enumerate}
\item
A heteroclinic front or shock \cite{vSH},
\begin{eqnarray}
& &
\left\lbrace
\matrix{
\displaystyle{
A=A_0 \left((k/2) (\tanh k \xi/2 + \varepsilon) \right)^{1/2}
e^{\displaystyle{i[\alpha \Log \cosh k \xi/2 + K \xi - \omega t] }},\
\varepsilon^2=1,
}
\hfill \cr
\displaystyle{
(-1/2+i \alpha) \left[i(c-2 p K) + 2 \varepsilon (-2+i \alpha) p k \right] p
 - A_0^2 q=0,
}
\hfill \cr
\displaystyle{
\frac{i \gamma- \omega}{p}=
\left(\frac{c}{2 p}\right)^2
-\left(K - \frac{c}{2 p} + \varepsilon (1-i \alpha)k/2 \right)^2.
}
\hfill \cr}
\right.
\label{eqCGL5Front}
\end{eqnarray}
The number of constraints among $(p,q,r,\gamma)$ is
either two  (case $s_i=0$, $c$ arbitrary),
or     zero (case $s_i \not=0$, with a fixed velocity).

\item
A homoclinic source or sink \cite{MCC1994},
\begin{eqnarray}
& &
\left\lbrace
\matrix{
\displaystyle{
A=A_0 \left(\frac{k \sinh k a}{\cosh k \xi + \cosh k a} + r_0 \right)^{1/2}
e^{\displaystyle{i[\alpha \Log(\cosh k \xi + \cosh k a) + K \xi - \omega t]}}
},
\hfill \cr
\hfill \cr
\displaystyle{
c - 2 p K=0,\
\hbox{ which implies } c p_i=0,
}
\hfill \cr
\displaystyle{
(-1/2+i \alpha) \left[-2 k\mu_0 (1-i \alpha) + 2 (-2+i \alpha) r_0 \right] p
 - A_0^2 q=0,
}
\hfill \cr
\displaystyle{
\frac{i \gamma- \omega}{p}=
\left(\frac{c}{2 p}\right)^2
+ (1/2-i \alpha)^2 k^2
+ (3 -10 i \alpha -4 \alpha^2) k \mu_0 r_0
+ (3-8 i \alpha - 2 \alpha^2) r_0^2/2,
}
\hfill \cr
\displaystyle{
r_0 \left(r_0^2 + 2 k \mu_0 r_0 + k^2\right)=0.
}
\hfill \cr
}
\right.
\label{eqCGL5source}
% c s_i=0,\
%\gamma=(3 d_r-d_i)\frac{-12(d_i-d_r)s_r +(5 d_r-7 d_i) s_i}{16 (s_r^2+s_i^2)}
%\label{eqCGL5sourceConstraint}
\end{eqnarray}
in which $K,k^2,r_0,\mu_0=\coth k a$ are real constants.
The number of constraints among $(p,q,r,\gamma)$ is
either two (case $s_i=0$, with $c$ arbitrary),
or one (case $s_i \not=0, c=0$).

\end{enumerate}

Each of these solutions depends on two additional signs arising from
the resolution of (\ref{eqCGL5LeadingOrderComplex})
\begin{eqnarray}
& &
A_0^2 =\varepsilon_2 \sqrt{\frac{2 e_r + \varepsilon_1 \Delta}{e_i^2}},\
\alpha=                    \frac{2 e_r + \varepsilon_1 \Delta}{2 e_i},\
\Delta=\sqrt{4 e_r^2 + 3 e_i^2},\
\varepsilon_1^2=\varepsilon_2^2=1.
\label{eqCGL5LeadingOrderReal}
\end{eqnarray}

% ========================================================================
\subsection{Swift-Hohenberg}
\label{sectionKnownSolutionsSH}

\index{Swift-Hohenberg equation}

Again, $A_0^4$ and $\alpha$ denote two real constants defined by
\begin{eqnarray}
& &
(-1+i \alpha) (-2+i \alpha)(-3+i \alpha) (-4+i \alpha) b + A_0^4 r=0.
\label{eqSHLeadingOrderComplex}
\end{eqnarray}
In the case $\Im(r/b) \not= 0$ to which we restrict ourselves,
only two solutions seem to be currently known.

\begin{enumerate}
\item
A stationary front \cite[Eq.~(127)]{MAA}
\begin{eqnarray}
& &
\left\lbrace
\matrix{
\displaystyle{
A=A_0 (k/2) \tanh k x/2\
e^{\displaystyle{i[\alpha \Log \cosh k x/2 - \omega t] }},
}
\hfill \cr
\displaystyle{
k^2=
\frac{2}{5(2-i \alpha) b}
\left[\frac{A_0^2 q}{(1-i \alpha)(2-i \alpha)}+p\right],
}
\hfill \cr
\displaystyle{
i \gamma- \omega=
\frac{16 -30 i \alpha -15 \alpha^2}{16} b k^4 + \frac{-2+3 i \alpha}{4} p k^2.
}
}
\right.
\label{eqSHFrontStationary}
\end{eqnarray}

\item
A stationary pulse \cite[Eq.~(119)]{MAA}
\begin{eqnarray}
& &
\left\lbrace
\matrix{
\displaystyle{
 A=A_0 (-i k \sech k x) % i?
 e^{\displaystyle{i[\alpha \Log \cosh k x - \omega t ]}},
}
\hfill \cr
\displaystyle{
k^2=\frac{-1}{2(5-4 i \alpha-\alpha^2)b}
\left[\frac{A_0^2 q}{(1-i \alpha)(2-i \alpha)}+p\right],
}
\hfill \cr
\displaystyle{
i \gamma- \omega=(1-i \alpha)^2 \left[(1-i \alpha)^2 b k^4 + p k^2\right].
}
}
\right.
\label{eqSHPulse}
\end{eqnarray}
In both solutions,
the number of constraints on $(b,p,q,r,\gamma)$ is two
(defined by the vanishing of the imaginary part of the relations
for $k^2$ and $i \gamma - \omega$).

\end{enumerate}

% ========================================================================
\section{Investigation of the amount of integrability}
\label{sectionCount}

% ========================================================================
\subsection{Counting arguments based on singularity analysis}
\label{sectionSingularityAnalysis}

Because the ODE (\ref{eqODEReduced}) is assumed nonintegrable,
the number of integration constants which can be present in any
closed form solution
is strictly smaller than the differential order of the ODE.
Let us first compute precisely this difference,
an indicator of the amount of integrability of the ODE.
The technique to do so is just the Painlev\'e test
(see Ref.~\cite{Cargese1996Conte} for the basic vocabulary of this technique).
Let us present it on the KS example (\ref{eqKSODE}).
\index{Kuramoto--Sivashinsky (KS)!equation}

Looking for a local algebraic behaviour
near a movable singularity $x_0$
(\textit{movable}
  \index{movable}
 means: which depends on the initial conditions),
% [NDLR How to put $x \to x_0$ under $\sim$?]
% \usepackage{amsmath}         A \xrightarrow[x\rightarrow a]{} B
\begin{eqnarray}
& &
u \sim_{x \to x_0} u_0 \chi^p,\ u_0 \not=0,\ \chi=x-x_0,
\end{eqnarray}
one first obtains the usual balancing conditions
(here, between the highest derivative and the nonlinearity)
\begin{eqnarray}
& &
p-3=2p,\ p(p-1)(p-2) \nu u_0 + \frac{u_0^2}{2}=0,\
\label{eqKSLeadingOrder}
\end{eqnarray}
easily solved as
\begin{eqnarray}
& &
p=-3,\ u_0=120 \nu,
\label{eqKSLeadingOrderSol}
\end{eqnarray}
which yields the Laurent series,
\begin{eqnarray}
& &
u^{(0)} = 120 \nu \chi^{-3} - 15b \chi^{-2}
        + \frac{15 (16 \mu \nu - b^2)}{4 \times 19 \nu} \chi^{-1}
        + \frac{13 (4  \mu \nu - b^2) b}{32 \times 19 \nu^2}
        + O(\chi^1),
\label{eqKSODELaurent}
\end{eqnarray}
from which two out of the three arbitrary constants are missing.
These two constants appear in perturbation \cite{CFP1993},
\begin{eqnarray}
& &
u= u^{(0)} + \varepsilon u^{(1)} + \varepsilon^2 u^{(2)} + \dots,
\end{eqnarray}
in which the small parameter $\varepsilon$ is not in the ODE (\ref{eqKSODE}).
The linearized equation around $u^{(0)}$
\begin{eqnarray}
& &
\left(\nu \frac{\D^3}{\D x^3} + b \frac{\D^2}{\D x^2} + \mu \frac{\D}{\D x}
 + u^{(0)}\right) u^{(1)}=0,
\end{eqnarray}
is then of the Fuchsian type near $x=x_0$,
with an indicial equation
($q=-6$ denotes the singularity degree of the \LHS\ $E$ of (\ref{eqKSODE}))
\begin{eqnarray}
& &
\lim_{\chi \to 0} \chi^{-j-q} (\nu \partial_x^3 + u_0 \chi^p) \chi^{j+p}
\\
& &
= \nu (j-3)(j-4)(j-5) + 120 \nu = \nu (j+1) (j^2 -13 j + 60)
\\
& &
=\nu (j+1) \left(j-\frac{13 + i \sqrt{71}}{2}\right)
           \left(j-\frac{13 - i \sqrt{71}}{2}\right)=0.
\end{eqnarray}
The local representation of the general solution,
\begin{eqnarray}
u(x_0,\varepsilon c_+,\varepsilon c_-)
& = &
 120\nu \chi^{-3} \{
 \hbox{Taylor}(\chi)
\nonumber
\\
& &
+ \varepsilon [
              c_{+} \chi^{(13+i\sqrt{71})/2} \hbox{Taylor}(\chi)
\nonumber
\\
& &
        + \ \ c_{-} \chi^{(13-i\sqrt{71})/2} \hbox{Taylor}(\chi) ]
            + {\mathcal O}(\varepsilon^2)\},
\nonumber
\end{eqnarray}
in which ``Taylor'' denotes converging series of $\chi$,
does depend on three arbitrary constants
$(x_0,\varepsilon c_+,\varepsilon c_-)$
(the Fuchs index $-1$ only represents a shift of $x_0$).
The dense movable branching arising from the two irrational indices
characterizes \cite{TF} the chaotic behaviour,
and the only way to remove it is to require
$\varepsilon c_+=\varepsilon c_-=0$, i.e.~$\varepsilon=0$,
thus restricting to a single arbitrary constant
the analytic part of the solution.

To summarize, let us introduce two notions.

The first one is trivial.
One calls
\textit{irrelevant}
 \index{irrelevant}
any integration constant which, because of some symmetry,
is always present in any solution.
The KS ODE has one such irrelevant integration constant,
the origin $\xi_0$ of $\xi$,
and we will systematically omit to write it.
The traveling wave reduction of CGL3 has two irrelevant integration constants,
the origins of $\xi$ and $\varphi$, etc.

The second notion is quite an important property of the equation.
We will call
\textit{unreachable}
 \index{unreachable}
any constant of integration which cannot participate to any
closed form solution.
The KS ODE has two unreachable integration constants.

We will also call
\textit{general analytic solution}
 \index{general analytic solution}
the closed form solution which depends on the maximal possible number
of reachable integration constants,
and our goal is precisely to exhibit a closed form expression
for this general analytic solution,
whose local representation is a Laurent series like
(\ref{eqKSODELaurent}).

The above notions (irrelevant, unreachable) are attached to an
equation, not to a solution.
Let us similarly introduce two integer numbers,
attached to a solution,
allowing one to quantify how far this solution is distant from
the general analytic solution.

We will call
\textit{deficiency} of a closed form solution
 \index{deficiency}
the number of reachable integration constants,
excluding the irrelevant ones,
which are missing in this solution.
In KS for instance,
the elliptic solution has a zero deficiency ($A$ is arbitrary),
and all the trigonometric solutions have deficiency one ($A$ is fixed).

Let us finally define the
\textit{codimension} of a closed form solution of an equation
 \index{codimension}
as the number of constraints on the fixed parameters
(\textit{fixed} \index{fixed} means: which occur in the definition
of the equation).
Thus, the elliptic and trigonometric
solutions of KS have codimension one,
and the rational solution has codimension two.

% ============================================================================
\subsection{Evidence for unknown solutions} 

Computer simulations as well as real experiments
(for a recent review, see \cite{vS2003})
sometimes display
regular patterns in the $(x,t)$ plane,
and some of them are indeed described by some analytic solution.
For the remaining patterns,
the guess is that there should exist analytic expressions, to be found,
corresponding to these patterns.

For the KS equation (\ref{eqKS}),
one thus observes a homoclinic solitary wave \cite[Fig.~7]{Toh}
$\varphi=f(\xi),\xi=x-ct$,
while all solutions known to date are heteroclinic.
\index{Kuramoto--Sivashinsky (KS)!equation}

For the CGL3 equation (\ref{eqCGL3}),
it has been predicted \cite{vanHecke}
the existence of a fourth physically interesting solution,
which is a codimension-one
homoclinic hole solution with an arbitrary velocity $c$.
\index{complex Ginzburg--Landau!(CGL3) equation}

Table \ref{TableCounting} gathers the current state of the known solutions
for the various nonintegrable equations considered in this chapter.
In Ref.~\cite{vHSvS},
another counting, based on the various possible topological structures,
is made for CGL3 and provides the same results.

\begin{table}[h] % [p]
\caption[garbage]{
Integer numbers rating 
the particular solutions of a nonintegrable equation.
The vocabulary (irrelevant, unreachable, deficiency, codimension)
is defined in section \ref{sectionSingularityAnalysis}.
The column ``Available''
indicates the number of relevant, reachable integration constants,
this is the algebraic sum
``Order'' -- ``Irrelevant'' -- ``Unreachable''.
The last two columns indicate the properties of the solutions
in their order of appearance section \ref{sectionKnownSolutions}.
The best solution would be one with deficiency and codimension both
equal to zero, with the reduction parameters $c$ and $\omega$ arbitrary.
% ? NDLR Achtung! Do not confuse PDE and ODE(x-ct)
}
\vspace{0.2truecm}
\begin{center}
\begin{tabular}{| c | c | l | c | c || l | l |}
%\hline % \hline % ********************************************************
\hline % \hline % ********************************************************
Equation & Order & Irrelevant & Unreachable & Available
         & Deficiency & Codimension
\\ \hline \hline % ********************************************************
  CGL3 & 4 & $2=\xi_0,\varphi_0$ & 2 & 0 & 0,0,0       & 1,2,2
\\ \hline % \hline % ********************************************************
  CGL5 & 4 & $2=\xi_0,\varphi_0$ & 2 & 0 & 0,0         & 1,1
\\ \hline % \hline % ********************************************************
  KS   & 4 & $1=\xi_0$           & 2 & 1 & 0,1,1,1,1,1,1,1 & 1,1,1,1,1,1,1,2
\\ \hline % \hline % ********************************************************
  SH   & 8 & $2=\xi_0,\varphi_0$ & 6 & 0 & 0,0         & 2,2
\\ \hline % \hline % ********************************************************
\end{tabular}
\end{center}
\label{TableCounting}
\end{table}

% ========================================================================
\section{Selection of possibly single valued dependent variables}
\label{sectionSelectionSinglevalued}

Whatever be the class of methods to be applied,
a prerequisite is
to determine a variable whose dominant behaviour is single valued
and which satisfies some algebraic ODE (or more generally PDE).
This is the case for KS since the solution $p$ of (\ref{eqKSLeadingOrder})
is integer,
but in CGL3, CGL5, or SH this is the case of neither $(A,\bar A)$,
nor $\arg A$,
and for CGL5 this is not even the case of $\mod{A}$.
Indeed, considering CGL3 for instance,
the dominant terms are
 \index{dominant terms}
\index{complex Ginzburg--Landau!(CGL3) equation}
\begin{eqnarray}
& &
p A_{xx} + q \mod{A}^2 A,\
\end{eqnarray}
and one easily checks that $\mod{A}$ generically behaves like simple poles.
Let us therefore define the dominant behaviour of the two fields
$(A, \overline{A})$ as
\begin{eqnarray}
& &
           \GLA  \sim           A_0 \chi^{-1+i \alpha},\
 \overline{\GLA} \sim \overline{A_0}\chi^{-1-i \alpha},\
A_0 \in \mathcal{C},\
\alpha \in \mathcal{R},
\end{eqnarray}
in which $(A_0,\alpha)$ are constants to be determined.
The resulting complex equation (equivalent to two real equations)
\begin{eqnarray}
& & (-1+i \alpha)(-2+i \alpha) p + \mod{A_0}^2 q =0,\
\label{eqCGL3Leading}
\end{eqnarray}
is precisely the one artificially introduced earlier as
(\ref{eqCGL3LeadingOrderComplex})
and solved in (\ref{eqCGL3LeadingOrderReal}),
the convention that $A_0$ is real being allowed by the phase invariance
of CGL3.
The same applies to CGL5
\begin{eqnarray}
& &
           \GLA  \sim           A_0 \chi^{-1/2+i \alpha},\
 \overline{\GLA} \sim           A_0 \chi^{-1/2-i \alpha},\
\end{eqnarray}
see (\ref{eqCGL5LeadingOrderComplex}) and (\ref{eqCGL5LeadingOrderReal}),
and to SH,
\index{Swift-Hohenberg equation}
\begin{eqnarray}
& &
           \GLA  \sim           A_0 \chi^{-1+i \alpha},\
 \overline{\GLA} \sim           A_0 \chi^{-1-i \alpha},\
\end{eqnarray}
see (\ref{eqSHLeadingOrderComplex}).

In all three examples CGL3, CGL5, SH,
the variable $M=\mod{A}^2$
satisfies an algebraic ODE, which can be built by elimination of $\arg A$,
and it has a single valued dominant behaviour
(respectively movable double poles, simple poles, double poles).
Moreover, again for CGL3, CGL5, SH,
for all the solitary wave solutions which are known to date
(see section \ref{sectionKnownSolutions}),
this variable $M$ is represented by quite simple mathematical expressions,
namely
either polynomials in one elementary variable $\tau$, Eq.~(\ref{eqtau}),
which satisfies a Riccati equation
\begin{eqnarray}
& &
\frac{\D}{\D z} \tau(z)=1-\tau^2,\
\tau=\tanh z,
\label{eqRiccatitau}
\end{eqnarray}
or (source solution of CGL5, Eq.~(\ref{eqCGL5source}))
polynomials in two elementary variables $(\sigma,\tau)$
which satisfy a projective Riccati system \cite{CM1993}
\begin{eqnarray}
& &
\frac{\D}{\D z}\tau=1-\tau^2- \mu_0 \sigma,\
\frac{\D}{\D z}\sigma=- \sigma \tau,\
\sigma^2-\tau^2 - 2 \mu_0 \sigma + 1=0,
\label{eqProjectiveRiccati}
\end{eqnarray}
in which $\mu_0$ is a constant,
and whose solution can be expressed as
\begin{eqnarray}
& &
\tau=\frac{\sinh z}{\cosh z + \cosh k a},\
\sigma=\frac{\sinh k a}{\cosh z + \cosh k a},\
\mu_0=\coth k a.
\label{eqtausigma}
\end{eqnarray}
When $\mu_0 (\mu_0^2-1)=0$, the class of polynomials in $(\sigma,\tau)$
degenerates to the class of polynomials in $(\sech,\tanh)$.

Therefore
$M=\mod{A}^2$ will be our best choice to search for closed form solutions
of CGL3, CGL5, SH.

\textit{Remark}.
Despite the multivalued dominant behaviour of the complex amplitude $A$
of CGL3 and SH,
one can define two variables with a single valued dominant behaviour.
In this complex modulus representation \cite{CM1993}
\index{complex modulus}
\begin{eqnarray}
& &
          A =A_0           Z (\xi) e^{\displaystyle{ i[\Phi(\xi)-\omega t]}},\
\overline{A}=A_0 \overline{Z}(\xi) e^{\displaystyle{-i[\Phi(\xi)-\omega t]}},\
\label{eqComplexModulus}
\end{eqnarray}
with $Z$ complex and $\Phi$ real,
the dominant behaviour is
\begin{eqnarray}
           Z  \sim        \chi^{-1},\
 \overline{Z} \sim        \chi^{-1},\
 \Phi'        \sim \alpha \chi^{-1},
\end{eqnarray}
and the truncation of $(Z,\overline{Z},\Phi')$
might prove to be much more economical than that of $M$.
All the solutions listed in section \ref{sectionKnownSolutions}
for CGL3, CGL5, SH have been written in this representation.

% ========================================================================
\section{On the price to obtain closed form expressions}
\label{sectionBothClasses}

Let us now give some details on the distinction
between the two main classes of methods outlined in the introduction.

In the \textbf{first class of methods},
one gives as an input some class of expressions $f(\xi)$
(for instance polynomials in $\sech k \xi$ and $\tanh k \xi$),
and by a direct computation one checks whether there indeed are some
solutions in the given class.
We will call for shortness these methods \textit{sufficient},
because they for sure miss any solution outside the given class, e.g.
for the ODE
\begin{eqnarray}
& &
{M'}^2+\left(12 M^2 - \frac{3}{2}\right) M'
 + 36 M^4 - \frac{17}{2} M^2 + \frac{1}{2}=0,
\end{eqnarray}
its solution since it is rational in $\tanh k \xi$,
\begin{eqnarray}
& &
M=\frac{\tanh(\xi-\xi_0)}{2+\tanh^2(\xi-\xi_0)}.
\label{eqrationaltanh}
\end{eqnarray}

In the \textbf{second class of methods},
the search for first order autonomous subequations (\ref{eqOrder1Autonomous})
requires no \textit{a priori} assumption at all,
and, from the classical results recalled in Appendix,
the knowledge of the first order subequation is indeed equivalent
to the knowledge of the explicit expression (\ref{eqFormalSolution}).
As opposed to the previous methods,
which are ``sufficient'' as said above,
the proposed method can be qualified as ``necessary''.

The difference between the two classes of methods is obvious:
the class of expressions $f(\xi)$ is an output of the second method,
while it is an input of the first one.
This is why the second method can find, if they exist,
not only some but \textit{all} the solutions which are elliptic
or trigonometric.

\textit{Remark}.
The cost of the method of first order autonomous subequations
is an increasing function of the positive integer $m$ occuring in
(\ref{eqsubeqODEOrderOnePP}),
but $m$, which is an input of the method, is not bounded.
Indeed, any rational function
$u=P_N(\tanh(\xi-\xi_0))/P_D(\tanh(\xi-\xi_0))$
satisfies an ODE (\ref{eqOrder1Autonomous})
of order one and degree $\max(N,D)$.
By considering only some differential consequence of this ODE,
one cannot guess the correct value of $m$ in advance.

% ========================================================================
\section{First class of methods: truncations}
\label{sectionFirstClass}

After having selected, as
indicated in section \ref{sectionSelectionSinglevalued},
dependent variables with a single valued leading behaviour,
the methods called truncations consist in defining for each such
dependent variable some single valued closed form class of expressions,
then in checking whether there exist solutions in that class.

The class of expressions to choose as an input depends on the number of
\textit{families of movable singularities}
   \index{family of movable singularities}
of the considered dependent variable.
Thus, the field $u$ of KS has only one family,
i.e.~one value of $u_0$,
while the field $M=\mod{A}^2$ of CGL3, CGL5 or SH
has respectively two, four, and four families.
Let us start with the simplest class of expressions.

% ========================================================================
\subsection{Polynomials in $\tanh$ (one-family truncation)}
\label{sectionOneFamilyTruncation}

The class of polynomials in $\tanh (k/2)\xi$ is the most frequently
encountered class of closed form solutions of autonomous PDEs.
This fact is the direct consequence of a quite remarkable property.
Indeed, by a result of Painlev\'e,
the variable $\tau$ in (\ref{eqRiccatitau}) is the unique variable
to be at the same time single valued and closed by differentiation:
if $u$ is such a polynomial,
\begin{eqnarray}
& &
u=\sum_{j=0}^{-p} u_j \chi^{j+p},\
\chi^{-1}=\frac{k}{2} \tanh\frac{k}{2}\xi,\ \xi=x-ct,
\label{eqTruncationuOneFamily}
\end{eqnarray}
the \LHS\ $E(u,x,t)$ of the equation of the PDE
is also such a polynomial,
\begin{eqnarray}
& &
E=\sum_{j=0}^{-q} E_j \chi^{j+q},\
\label{eqTruncationEOneFamily}
\end{eqnarray}
and its identification to the null polynomial
\begin{eqnarray}
& &
\forall j\ : E_j=0,\
\end{eqnarray}
generates the smallest possible number of
\textit{determining equations} $E_j=0$.
 \index{determining equations}
As compared to the Laurent series (\ref{eqKSODELaurent}),
the series (\ref{eqTruncationuOneFamily}) terminates,
hence its name of
\textit{truncation}.
 \index{truncation}

The truncation (\ref{eqTruncationuOneFamily}) involves only one
value of $u_0$,
it is called for this reason a
\textit{one-family truncation}.
 \index{truncation!one-family}
Let us give a few examples.

% ==========================================================================
\subsubsection{One-family truncation of the KS equation}
%\label{sectionKSOneFamilyTruncation}

\index{Kuramoto--Sivashinsky (KS)!equation}

The symbols $u_0$ and $p$ denoting the leading behaviour of the ODE
(\ref{eqKSODE}),
the truncation (\ref{eqTruncationuOneFamily}) defines $-q+1=7$
determining equations, the first four being
\begin{eqnarray}
& &
E_0 \equiv -60 \nu u_0 + \frac{u_0^2}{2}=0,\
\\
& &
E_1 \equiv 12 b u_0 + (u_0 -24 \nu) u_1=0,\
\\
& &
E_2 \equiv
 -3 \mu u_0 + \frac{57}{2} k^2 \nu u_0 + 6 b u_1 + \frac{1}{2} u_1^2
+(u_0 -6 \nu) u_2=0,\
\\
& &
E_3 \equiv 
- \frac{9}{2} b k^2 u_0 -2 \mu u_1 + 10 k^2 \nu u_1 + 2 b u_2 + u_1 u_2
 + u_0 u_3=0.
\end{eqnarray}
%in which $k^2=-2 S$.
The structure of this kind of algebraic determining equations
is always the same:
one algebraic equation for $u_0$ ($j=0$),
followed by $-p$ equations linear in $u_j,j=1, \cdots,-p$.
Equation $j=0$ has already been solved, see (\ref{eqKSLeadingOrderSol}),
and the next equations $j=1,\dots,-p$
have the same solution $u_j$ as in the infinite Laurent series
(\ref{eqKSODELaurent}).
The truncated expansion (\ref{eqTruncationuOneFamily}) then evaluates to
\begin{eqnarray}
& &
u=\mathcal{D} \Log \psi + \hbox{constant},
\label{eqOneFamilyTruncationWithD}
\end{eqnarray}
in which $\mathcal{D}$ is the singular part operator
\index{singular part operator}
defined in (\ref{eqKSD}) from the Laurent series,
and $\psi$ is the logarithmic primitive of $\chi^{-1}$,
an entire function defined by
\begin{eqnarray}
& &
\psi'' - \frac{k^2}{4} \psi =0,
\label{eqOrder2EntireTrigo}
\end{eqnarray}
whose value can be chosen without loss of generality as
\begin{eqnarray}
& &
\psi=\cosh \frac{k}{2} \xi.
\end{eqnarray}

After the operator $\mathcal{D}$ has been computed,
the two equations
(\ref{eqOneFamilyTruncationWithD}), (\ref{eqOrder2EntireTrigo})
are an equivalent way of defining a one-family truncation,
much more elegant than with
(\ref{eqTruncationuOneFamily}), (\ref{eqRiccatitau}).

The remaining $-q+p$ equations $j=-p+1,\dots,-q$ are algebraic
in $k^2$ and the parameters appearing in the definition of the equation 
(\ref{eqKSODE}) (one says the \textit{fixed} parameters),
\begin{eqnarray}
& &
E_4 \equiv 
- \frac{5}{2} b^2 k^2 + \frac{44}{19} \mu^2 + \frac{131}{304} b^4 \nu^{-2}
-\frac{87}{38} b^2 \mu \nu^{-1} + 40 k^2 \mu \nu - 76 k^4 \nu^2=0,\
\\
& &
E_5 \equiv b (b^2-16 \mu \nu) 
\left(5 k^2 +\frac{13}{152} b^2 \nu^{-2} - \frac{7}{19} \mu\nu^{-1}\right)=0,\
\\
& &
E_6 \equiv 
32 A + 3 \nu u_0 k^6 + 4 (b u_1  - \nu u_2) k^4 + 8 k^2 \mu u_2 + 16 u_3^2=0,
\end{eqnarray}
and they admit only the six solutions listed in Table \ref{TableKS}.

% ==========================================================================
\subsubsection
 {One-family truncation of the real modulus of CGL3}
\label{sectionCGL3OneFamilyTruncation}

\index{complex Ginzburg--Landau!(CGL3) equation}
Whatever be the chosen representation
(couple $(M,\varphi)$, $(Z,\overline{Z},\Phi)$, etc),
the CGL3 equation has more than one family,
see (\ref{eqCGL3LeadingOrderReal}),
therefore any one-family truncation only captures part of the whole
singularity structure and cannot yield the general analytic solution.
\index{general analytic solution}
Nevertheless, as already noticed
in section \ref{sectionKnownSolutionsCGL3} Eq.~(\ref{eqtau}),
the one-family truncation of $M=\mod{A}^2$ must provide at least
the three currently known solutions.
Let us perform it.

The field $M$
has two families of movable double pole-like singularities
\begin{eqnarray}
& &
M = \frac{3(3 d_r \pm  \Delta)}{2 d_i^2} \chi^{-2}
\left(1 + \frac{c s_i}{3} \chi + O(\chi^2)\right),\
\label{eqCGL3mod2Laurent}
\end{eqnarray}
with singular part operators $\mathcal{D}_\pm$ equal to
\begin{eqnarray}
\mathcal{D}_\pm=\frac{3(3 d_r \pm \Delta)}{2 d_i^2}
\left(- \partial_x^2 + \frac{c s_i}{3} \partial_x\right).
\label{eqCGL3OperatorD}
\label{eqCGL3D}
\end{eqnarray}

In its elegant definition, the one-family truncation,
\begin{eqnarray}
& &
\left\lbrace
\begin{array}{ll}
\displaystyle{
M=\mathcal{D}_\pm \Log \psi + m,
}
\\
\displaystyle{
\psi'' + \frac{S}{2} \psi=0,\ S=-\frac{k^2}{2}= \hbox{ constant},
}
\end{array}
\right.
\label{eqCGL3mod2Trunc}
\end{eqnarray}
transforms (\ref{eqCGL3Order3}) into the truncated Laurent series
\begin{eqnarray}
& &
\sum_{j=0}^{14} E_j \chi^{j-14}=0,
\end{eqnarray}
and one must solve the $15$ real determining equations $E_j=0$ for
the two constant unknowns $(S,m)$ and 
the five parameters $d_r,d_i,g_r,g_i,c s_i$
occurring in (\ref{eqCGL3Order3}).
By construction of $\mathcal{D}_{\pm}$,
equations $E_j=0, j=0,1,$ are identically zero.

To avoid carrying heavy expressions,
%(mainly undesirable denominators),
let us make the following nonrestrictive simplification.
Out of the five % fixed
parameters $d_r,d_i,g_r,g_i,c s_i$
of the ODE (\ref{eqCGL3Order3}),
only three are essential
($g_r,g_i,c$, equivalent to $\gamma,\omega,c$).
Indeed, $p$ and $q$ (i.e. $d_r+i d_i$ and $s_r - i s_i$)
can be rescaled to convenient numerical values,
%making rational all the coefficients of the above solutions,
such as
\begin{eqnarray}
{\hskip -10.0 truemm}
& &
p=-1-3 i,\
q= 4-3 i,\
\nonumber
\\
& &
d_r= \frac{1}{2},\
d_i= \frac{3}{2},\
s_r=-\frac{1}{10},\
s_i=-\frac{3}{10},\
\Delta=\frac{9}{2}.
\label{eqCGL3Num}
\end{eqnarray}

Choosing the $+$ sign in (\ref{eqCGL3mod2Trunc}),
one has $\mathcal{D}_{+} = 4 (-\partial_x^2 -(c/10) \partial_x)$.
As seen from the first few determining equations,
\begin{eqnarray}
& &
E_2 \equiv 
\frac{57}{100} c^2 + 156 c_2 + 13 k^2 + 4 g_i + 16 g_r =0,\
\\
& &
E_3 \equiv 
\left(- \frac{39}{25} c^2 -432 c_2 -28 k^2 -16 g_i -48 g_r \right) c =0,\
\end{eqnarray}
the resolution presents no difficulty.
In particular,
after solving the equations numbered $j=0,\dots,6$,
all the remaining equations are identically zero,
a fact which indicates a high redundancy in these determining equations,
which are therefore not at all optimal.
In the CGL3 case properly said $\Im(p/q)\not=0$,
for each sign in (\ref{eqCGL3D}) one obtains three solutions,
\begin{eqnarray}
& &
\left\lbrace
\matrix{
\displaystyle{
M= -2 \left\lbrack\left(\tau-\frac{c}{20}\right)^2+\left(\frac{c}{10}\right)^2
      \right\rbrack,\
\varphi'-\frac{c s_r}{2}=
- \tau - \frac{c}{20} - \frac{c}{5 M} \left(\tau^2 - \frac{k^2}{4}\right),\
}
\hfill \cr
\displaystyle{
k^2=- 7 \left(\frac{c}{10}\right)^2 - \frac{4}{3} g_r,\
3 g_i + 2 g_r + \frac{3 c^2}{50}=0,\
}
\hfill \cr}
\right.
\label{eqCGL3Hole-}
% ************************************
\\
& &
\left\lbrace
\matrix{
\displaystyle{
M= -2 \left(\tau^2 - \frac{k^2}{4}\right),\
\varphi'-\frac{c s_r}{2}=- \tau,
}
\hfill \cr
\displaystyle{
k^2=2 g_r,\
c=0,\ g_i=0,\
}
\hfill \cr}
\right.
\label{eqCGL3Pulse-}
% ************************************
\\
& &
\left\lbrace
\matrix{
\displaystyle{
M= -2 \left(\tau \pm \frac{k}{2}\right)^2,\
\varphi'-\frac{c s_r}{2}= - \tau + \frac{c}{20},
}
\hfill \cr
\displaystyle{
k^2=\left(\frac{c}{10}\right)^2,\
g_r=0,\ g_i - \frac{c^2}{50}=0,\
}
\hfill \cr}
\right.
\label{eqCGL3Front-}
% ************************************
\end{eqnarray}
and
\begin{eqnarray}
& &
\left\lbrace
\matrix{
\displaystyle{
M= 4 \left\lbrack\left(\tau - \frac{c}{20}\right)^2
  + \left(\frac{c}{20}\right)^2
      \right\rbrack,\
\varphi'-\frac{c s_r}{2}=
2 \tau - \frac{c}{20} - \frac{c}{5 M} \left(\tau^2 - \frac{k^2}{4}\right),\
}
\hfill \cr
\displaystyle{
k^2=- \left(\frac{c}{10}\right)^2 + \frac{2}{3} g_r,\
3 g_i - g_r + \frac{3 c^2}{80}=0,\
}
\hfill \cr}
\right.
\label{eqCGL3Hole+}
% ************************************
\\
& &
\left\lbrace
\matrix{
\displaystyle{
M= 4 \tau^2,\
\varphi'-\frac{c s_r}{2}=2 \tau,
}
\hfill \cr
\displaystyle{
k^2=\frac{2}{3} g_r,\
c=0,\ 3 g_i - g_r =0,\
}
\hfill \cr}
\right.
\label{eqCGL3Pulse+}
% ************************************
\\
& &
\left\lbrace
\matrix{
\displaystyle{
M= 4 \left(\tau \pm \frac{k}{2}\right)^2,\
\varphi'-\frac{c s_r}{2}=2 \tau - \frac{c}{10},
}
\hfill \cr
\displaystyle{
k^2=\left(\frac{c}{10}\right)^2,\
g_r=0,\ g_i - \frac{c^2}{50}=0.
}
\hfill \cr}
\right.
\label{eqCGL3Front+}
% ************************************
\end{eqnarray}
These solutions are identical to those listed, in the same order,
in section \ref{sectionKnownSolutionsCGL3}.

% ==========================================================================
\subsubsection
 {One-family truncation of CGL3 in the complex modulus representation}
\label{sectionCGL3OneFamilyTruncationRealModulus}

\index{complex Ginzburg--Landau!(CGL3) equation}
As already outlined at the end of section
\ref{sectionSelectionSinglevalued},
the one-family truncation of $(Z,\overline{Z},\Phi')$
\begin{eqnarray}
& &
\left\lbrace
\begin{array}{ll}
\displaystyle{
Z=\chi^{-1}+X+iY,
}
\\
\displaystyle{
\overline{Z}=\chi^{-1}+X-iY,
}
\\
\displaystyle{
\Phi= \alpha \Log \psi + K \xi,
}
\end{array}
\right.
\label{eqCGL3TruncZPhi}
\end{eqnarray}
with the gradient definitions 
\begin{eqnarray}
& &
\left\lbrace
\begin{array}{ll}
\displaystyle{
(\Log \psi)'=\chi^{-1},
}
\\
\displaystyle{
\chi'=1-\frac{k^2}{4} \chi^2,
}
\end{array}
\right.
\end{eqnarray}
puts the \LHS\ of Eq.~(\ref{eqCGL3}) in the form
\begin{eqnarray}
& &
\sum_{j=0}^{3} E_j \chi^{j-3}=0,
\end{eqnarray}
thus generating four complex determining equations $E_j=0$,
(i.e.~eight real, to be compared with the fifteen of section
\ref{sectionCGL3OneFamilyTruncation}).
These equations must first be solved as a \textit{linear} system
on $\mathcal{C}$, as follows \cite[Appendix A]{CM1993}.
The first equation $E_0=0$, identical to (\ref{eqCGL3LeadingOrderComplex}),
is linear in $p$ and $q$, let us solve it for $q$,
\begin{eqnarray}
& &
q=- (1-i \alpha) (2-i \alpha) A_0^{-2} p.
\label{eqCGL3Cpivot0}
\end{eqnarray}
The next equation $j=1$ is then linear in $K,X,Y,c$, 
let us solve it for instance for $K$,
\begin{eqnarray}
& &
K=(3 i + \alpha) X - Y + \frac{c}{2 p}.
\end{eqnarray}
The equation $j=2$, linear in $\gamma,\omega,k^2$,
is solved for $(i \gamma - \omega)/p$
\begin{eqnarray}
& &
\frac{i \gamma - \omega}{p} =
\left(\frac{c}{2p}\right)^2
+ [X-(1-i \alpha) i Y]^2
-(2-3 i \alpha) \left[\frac{k^2}{4}-(X+i Y)^2\right],
\end{eqnarray}
and the advantage of this pivoting elimination is that the last equation
$j=3$,
which does not depend on $(q,K,\gamma)$ by construction,
is also independent of $(p,c,\omega,A_0)$.
It only depends on $(X,Y,\alpha,k^2)$,
and it factorizes as
\begin{eqnarray}
& &
E_3 \equiv (2 X - \alpha Y) (4 (X+i Y)^2 - k^2)=0,
\label{eqCGL3Cpivot3}
\end{eqnarray}
thus defining two solutions on $\mathcal{C}$.

Finally, considering now the system
(\ref{eqCGL3Cpivot0})--(\ref{eqCGL3Cpivot3})
for the real unknowns or parameters
$(A_0^2,\alpha,K,c,X,Y,\gamma,\omega,k^2)$,
it is quite easy to obtain the three solutions listed 
in section \ref{sectionKnownSolutionsCGL3}.

% ========================================================================
\subsection{Polynomials in $\tanh$ and $\sech$ (two-family truncation)}
\label{sectionTwoFamilyTruncation}

\index{truncation!two-family}

The class of polynomials in $\tanh$ and $\sech$
\begin{eqnarray}
& &
u=\left( \sum_{j=0}^{-p}   a_j \tanh k \xi \right)
+ \left( \sum_{j=0}^{-p-1} b_j \tanh k \xi \right) \sech k \xi,
\end{eqnarray}
can equivalently be represented by the class of powers of $\tanh$
ranging from $p$ to $-p$ \cite{Pickering1993},
\begin{eqnarray}
& &
u=\sum_{j=0}^{- 2 p} u_j \chi^{j+p},\
\chi^{-1}=\frac{k}{2} \tanh\frac{k}{2}\xi,\ \xi=x-ct,\
u_0 u_{-2p}\not=0,
%\label{eqTruncationTwoFamilychi}
\end{eqnarray}
because of the elementary identities \cite{CM1993}
\begin{equation} 
   \tanh z - {1 \over \tanh z}= -2 i \sech\left[2 z + i {\pi \over 2}\right],\
   \tanh z + {1 \over \tanh z}=  2   \tanh\left[2 z + i {\pi \over 2}\right].
\label{eqIdentitiestanhsech}
\end{equation}

A solution in this class 
can only exist for ODEs admitting at least two families with the same $p$.
Indeed, if for this $p$ there exists only one value of $u_0$,
only the second combination $\tanh+1/ \tanh = 2 \tanh$ can contribute.
For instance, the KS equation cannot admit such a solution.
\index{Kuramoto--Sivashinsky (KS)!equation}

More generally, 
the class of polynomials in $\tau$ and $\sigma$ defined in
(\ref{eqtausigma}),
\begin{eqnarray}
& &
u=\left( \sum_{j=0}^{-p}   a_j \tau^j  \right)
+ \left( \sum_{j=0}^{-p-1} b_j \tau^j  \right) \sigma,\
(a_{-p},b_{-p-1})\not=(0,0),
\label{eqTruncationTwoFamilytausigma}
\end{eqnarray}
is equivalently defined as \cite[Appendix A]{CM1993} 
\begin{eqnarray}
& &
\left\lbrace
\begin{array}{ll}
\displaystyle{
u=\mathcal{D}_{1} \Log \psi_1 + \mathcal{D}_{2} \Log \psi_2 + m,\
m=\hbox{const},
}
\\
\displaystyle{
    \psi_1'' + \frac{S}{2} \psi_1=0,\
    \psi_2'' + \frac{S}{2} \psi_2=0,\
    S=-\frac{k^2}{2}= \hbox{ constant},
}
\\
\displaystyle{
\frac{\psi_1'}{\psi_1} \frac{\psi_2'}{\psi_2}
=
- \frac{S}{2} - \frac{k}{2} \mu_0
\left(\frac{\psi_1'}{\psi_1} - \frac{\psi_2'}{\psi_2}\right).
}
\end{array}
\right.
\label{eqTruncationTwoFamilypsi1psi2}
\end{eqnarray}
In this writing,
which is the natural extension of (\ref{eqCGL3mod2Trunc}) to two families,
the linear operators $\mathcal{D}_{1}$ and $\mathcal{D}_{2}$ 
are the singular part operators of two \textit{different} families,
\index{singular part operator}
the entire functions $\psi_1$ and $\psi_2$ 
obey the same second order linear equation,
but with a different choice of the integration constants,
\begin{eqnarray}
& &
\psi_1=\cosh \frac{k}{2} (\xi+a),\
\psi_2=\cosh \frac{k}{2} (\xi-a),\
\mu_0=\coth k a.
\end{eqnarray}
The case $\mu_0 (\mu_0^2-1)=0$ reduces to the class 
of polynomials in $\tanh$ and $\sech$.

The practical implementation is the following.
\begin{enumerate}
\item
For the class of polynomials in $\tanh$ and $\sech$,
one puts the \LHS\ $E(u)$ of the nonlinear ODE under the same form as $u$,
\begin{eqnarray}
& &
\left\lbrace
\begin{array}{ll}
\displaystyle{
u=\sum_{j=0}^{- 2 p} u_j \chi^{j+p},\
u_0 u_{-2p}\not=0,\
}
\\
\displaystyle{
\chi'=1+\frac{S}{2} \chi^2,\
S=-\frac{k^2}{2},
}
\\
\displaystyle{
E=\sum_{j=0}^{- 2 q} E_j \chi^{j+q},\
}
\\
\displaystyle{
\forall j:\ E_j=0.
}
\end{array}
\right.
\label{eqTruncationTwoFamilychi}
\end{eqnarray}
and one solves the set of $-2q+1$ determining equations $E_j=0$.
\begin{eqnarray}
& &
\end{eqnarray}

\item
For the class of polynomials in $\tau$ and $\sigma$ defined in
(\ref{eqtausigma}),
under the assumption (\ref{eqTruncationTwoFamilypsi1psi2}),
the \LHS\ $E(u)$ is first expressed as a polynomial of the two variables
$\psi_j'/ \psi_j,j=1,2$
\begin{eqnarray}
& &
\sum_{k=0}^{-q} \sum_{l=0}^{-q-k} E_{k,l}
\left(\frac{\psi_1'}{\psi_1}\right)^k \left(\frac{\psi_2'}{\psi_2}\right)^l
=0,
\label{eqTruncationTwoFamilypsi1psi2both}
\end{eqnarray}
which further reduces,
thanks to the third line of (\ref{eqTruncationTwoFamilypsi1psi2}),
to the sum of two polynomials of one variable,
\begin{eqnarray}
& &
E_0
+\left(\sum_{j=1}^{-q} E_j^{(1)} \left(\frac{\psi_1'}{\psi_1}\right)^j\right)
+\left(\sum_{j=1}^{-q} E_j^{(2)} \left(\frac{\psi_2'}{\psi_2}\right)^j\right)
=0.
\label{eqTruncationTwoFamilypsi1psi2split}
\end{eqnarray}
One then requires the vanishing of the $-2q+1$ determining equations
 \index{determining equations}
\begin{eqnarray}
& &
E_0=0,\ E_j^{(1)}=0,\ E_j^{(2)}=0,\ j=1,\dots,-q.
\end{eqnarray}

\end{enumerate}

As an example, let us apply this to the ODE
\begin{eqnarray}
& &
E(u) \equiv 
\left(\frac{\D u} {\D \xi}\right)^2 - \alpha^2 (u^2 - b^2)^2 + c=0.
\end{eqnarray}
It admits two families,
with singular part operators
$\mathcal{D}_{1}= \alpha^{-1} \partial_\xi,
 \mathcal{D}_{2}=-\alpha^{-1} \partial_\xi$.
The relation $\mathcal{D}_{2}=-\mathcal{D}_{1}$
implies 
$E_j^{(1)} + (-1)^j E_j^{(2)} \equiv 0,\ j=1,2,3,4$,
and only 5 out of the 9 determining equations are linearly independent.
Moreover, by construction of the singular part operators,
the two equations $j=4$ are identically satisfied.
The next equation $j=3$
\begin{eqnarray}
& &
E_3^{(1)}\equiv -2 \alpha^{-2} k \mu_0 - 4 \alpha^{-1} m=0,
\end{eqnarray}
is solved for $m$.
Then, the equation $j=2$
\begin{eqnarray}
& &
E_2^{(1)}\equiv 2 b^2 +\alpha^{-2} k^2 -\frac{3}{2} \alpha^{-2} (k \mu_0)^2=0,
\end{eqnarray}
is solved for $k^2$,
considering $k \mu_0$ as a single variable.
The remaining system
\begin{eqnarray}
& &
E_1^{(1)}\equiv \left(- 2 b^2 + \frac{1}{2} \alpha^{-2} (k \mu_0)^2 \right)
 (k \mu_0)=0,
\\
& &
E_0 \equiv
 - \alpha^2 b^4 + c + \frac{1}{2} b^2 (k \mu_0)^2
 - \frac{1}{16} 
\alpha^{-2} (k \mu_0)^4=0,
\end{eqnarray}
admits two solutions.
The first one $c=0,(k \mu_0)^2=(2 \alpha b)^2$ 
corresponds to a factorization of the equation $E(u)=0$ into
two Riccati equations and therefore must be rejected.
The second one
\begin{eqnarray}
& &
k \mu_0=0,\
\alpha^{2} b^4 -c=0,
\end{eqnarray}
defines a solution,
provided the indicated constraint on the fixed parameters $(\alpha,b,c)$ is
satisfied.
This solution
\begin{eqnarray}
& &
\mu_0=0,\
m=0,\
k^2=-2 (\alpha b)^2,\
u=\alpha^{-1} \frac{\D} {\D \xi} \Log 
\frac{\cosh (k/2) (\xi+a)}{\cosh(k/2)(\xi-a)}
\end{eqnarray}
is nothing else than $u= i (k/ \alpha) \sech k \xi$,
using the relation $\mu_0=\coth k a$.

Indeed, as opposed to the function $\tanh$,
which satisfies an ODE admitting only one family of movable singularities
(the Riccati equation),
the function $\sech$ (or more generally its homographic transform $\sigma$)
satisfies a first order second degree ODE
\begin{eqnarray}
& &
{\sech'}^2 + \sech^4 - \sech^2=0,
\end{eqnarray}
which admits two families of movable simple poles with opposite residues
\begin{eqnarray}
& &
\sech (\xi - \xi_0) \sim \pm i (\xi - \xi_0)^{-1}.
\end{eqnarray}

% ==========================================================================
\subsubsection{Two-family truncation of the real modulus of CGL3}
\label{sectionCGL3TwoFamilyTruncation}

\index{complex Ginzburg--Landau!(CGL3) equation}

$M$ admits exactly two families, 
so its two-family truncation is quite appropriate.
The two singular part operators are,
for each family,
defined in (\ref{eqCGL3D}),
with $\mathcal{D}_{1}=\mathcal{D}_{+},\mathcal{D}_{2}=\mathcal{D}_{-}$.
The assumption (\ref{eqTruncationTwoFamilypsi1psi2}),
with $p=-2,q=-14$,
transforms (\ref{eqCGL3Order3}) into 
the sum (\ref{eqTruncationTwoFamilypsi1psi2split})
of two polynomials of one variable,
and the four equations $j=14,13$ are identically zero,
by definition of $\mathcal{D}_{\pm}$.

For $p,q,\gamma$ arbitrary, the resolution of the $25$ remaining
determining equation is impossible to carry out by hand.
But the hand computation becomes possible with the generic numerical values
(\ref{eqCGL3Num}).
First, the system $j=12$
\begin{eqnarray}
E_{12}^{(1)} & \equiv &
\frac{57}{5} c^2
- 780 m - 520 k^2 + 78 c k \mu_0 + 390 (k \mu_0)^2 + 80 g_i + 320 g_r=0,
\\
E_{12}^{(2)} & \equiv &
\frac{3}{5} c^2 
+ 120 m -40 k^2 - 24 c k \mu_0 + 120 (k \mu_0)^2 + 20 g_i - 40 g_r=0,
\end{eqnarray}
is solved as a linear system for $m$ and $k^2$.
The next system $j=11$
\begin{eqnarray}
E_{11}^{(1)}&  \equiv &
-\frac{39}{125} c^3 - \frac{254}{25} c^2 k \mu_0
+\frac{468}{5} c (k \mu_0)^2 - 312 (k \mu_0)^3 
\nonumber \\ & &
- \frac{156}{5} c g_i
- 168 k \mu_0 g_i - \frac{104}{5} c g_r 
- 48 k \mu_0 g_r=0,
\\
E_{11}^{(2)} & \equiv &
-\frac{177}{500} c^3 + \frac{218}{25} c^2 k \mu_0
- 39 c (k \mu_0)^2 + 156 (k \mu_0)^3 
\nonumber \\ & &
- \frac{87}{5} c g_i
+ 84 k \mu_0 g_i - 2 c g_r 
+ 24 k \mu_0 g_r=0,
\end{eqnarray}
is linear in $(g_r,g_i)$,
with a Jacobian $J=c (3 c - 5 k \mu_0)$.
For the first subcase $J\not=0$,
after solving for $(g_r,g_i)$ as functions of $(c, k \mu_0)$,
the next system $j=10$ only depends on $k \mu_0/c$ and it admits no
solution.
The discussion of the second subcase $J=0$ leads to the conclusion,
only using the next system $j=10$, that no solution exists.
An identical result is achieved for arbitrary values of $(p,q,\gamma)$
using computer algebra.

This unfortunate situation is exceptional,
and only reflects the difficulty of CGL3.
Should such a solution exist, it would have the form
\begin{eqnarray}
& &
M=\left(\frac{\Delta}{2 d_i^2} \tanh + c_1\right) \sech
+  \frac{9 d_r}{2 d_i^2}  \tanh^2 + c_3 \tanh + c_4,
\label{eqCGL3TwoFamiliesSolution}
\end{eqnarray}
and the constraint $c_3=0$ could define a homoclinic hole solution,
just like the (yet analytically unknown) one of van Hecke \cite{vanHecke}.

% ==========================================================================
\subsubsection
 {Two-family truncation of CGL3 in the complex modulus representation}
\label{sectionCGL3OneFamilyTruncationComplexModulus}

\index{complex Ginzburg--Landau!(CGL3) equation}

Let us denote $(A_0,\alpha)$ and $(A_2,\alpha_2)$ two different
solutions of (\ref{eqCGL3LeadingOrderReal}).

In the complex modulus representation (\ref{eqComplexModulus}),
the two-family truncation of $(Z,\overline{Z},\Phi)$ is defined as
\cite[Appendix A]{CM1993},
\begin{eqnarray}
& &
\left\lbrace
\begin{array}{ll}
\displaystyle{
A = \left(A_0 (\partial_{\xi} \Log \psi_1(\xi) +X+iY)
         +A_2  \partial_{\xi} \Log \psi_2(\xi)\right)
e^{i[-\omega t + \Phi(\xi)]},
}
\\
\displaystyle{
\overline{A} = 
    \left(A_0 (\partial_{\xi} \Log \psi_1(\xi) +X-iY)
         +A_2  \partial_{\xi} \Log \psi_2(\xi)\right)
e^{-i[-\omega t + \Phi(\xi)]},
}
\\
\displaystyle{
\Phi= \alpha \Log \psi_1 + \alpha_2 \Log \psi_2 + K \xi,
}
\end{array}
\right.
%\label{eqCGL3TruncZPhi}
\end{eqnarray}
with the definitions for the derivatives of $(\psi_1,\psi_2)$
given by the last two lines of (\ref{eqTruncationTwoFamilypsi1psi2}).
The \LHS\ of Eq.~(\ref{eqCGL3}) then takes the form
(\ref{eqTruncationTwoFamilypsi1psi2split}) with $q=-3$,
and one solves the seven complex determining equations 
as a linear system on $\mathcal{C}$,
similarly to what has been done in section
\ref{sectionCGL3OneFamilyTruncationRealModulus}.

From the two equations $j=3$,
\begin{eqnarray}
& &
E_3^{(1)} \equiv A_0 \left((1-i \alpha) (2-i \alpha) p + A_0^2 q\right)=0,
\\
& &
E_3^{(2)} \equiv A_2 \left((1-i \alpha_2) (2-i \alpha_2) p + A_2^2 q\right)=0,
\end{eqnarray}
and the two relations implied by (\ref{eqCGL3LeadingOrderReal}),
\begin{eqnarray}
& &
\alpha  =\frac{d_i}{3} A_0^2,\
\alpha_2=\frac{d_i}{3} A_2^2.
\end{eqnarray}
one first proves that the only possibility is $A_2=-A_0$,
therefore the two complex equations $j=3$ are solved as
\begin{eqnarray}
& &
A_2=-A_0,\
\alpha_2=\alpha,\
q=- (1-i \alpha) (2-i \alpha) A_0^{-2} p.
\end{eqnarray}
At the level $j=2$,
the symmetric combination
\begin{eqnarray}
& &
E_2^{(1)}+E_2^{(2)}
 \equiv 
p(1-i \alpha) 
\left[(i \alpha-3)X -i Y +\left(i \alpha-\frac{3}{2}\right) k \mu_0\right]=0,
\end{eqnarray}
is solved for the two pieces of information
\begin{eqnarray}
& &
X=-\frac{1}{2} k \mu_0,\
Y=\frac{1}{2} \alpha k \mu_0,
\end{eqnarray}
then the antisymmetric combination
\begin{eqnarray}
& &
E_2^{(1)}-E_2^{(2)}
 \equiv 
p(1-i \alpha) \left[\frac{c}{2p}-K\right]=0,
\end{eqnarray}
is solved as
\begin{eqnarray}
& &
K=\frac{c}{2p}.
\end{eqnarray}
At the level $j=1$,
the symmetric combination is identically zero,
and
the antisymmetric combination is solved for $(i \gamma - \omega)/p$
(we omit the expression).
The remaining equation
\begin{eqnarray}
& &
E_0
 \equiv 
\mu_0 
\left[(2+(\alpha^2-2) \mu_0^2)+i \alpha (-4+(\alpha^2 +4) \mu_0^2) \right]=0
\end{eqnarray}
admits as only solution $\mu_0=0$.
Therefore, one obtains the unique solution
\begin{eqnarray}
& &
A_2=-A_0,\
\alpha_2=\alpha,\
q=- (1-i \alpha) (2-i \alpha) A_0^{-2} p,
\\
& &
X=0,\
Y=0,\
\mu_0=0,\
K=\frac{c}{2 p},
\\
& &
\frac{i \gamma - \omega}{p} =
\left(\frac{c}{2p}\right)^2
+ (1-i \alpha)^2 k^2,
\end{eqnarray}
The value of $K$ implies $c p_i=0$,
and the case $c=0$ represents the homoclinic pulse (\ref{eqCGL3Pulse}).

% ========================================================================
\subsection{Polynomials in $\wp$ and $\wp'$}
\label{sectionEllipticTruncation}

A class of polynomial elliptic functions
can be defined for instance with the Weierstrass function $\wp$ 
and its derivative
\cite{KudryashovElliptic,Samsonov1994},
\begin{eqnarray}
& &
u=\left( \sum_{j=0}^{[-p/2]}   a_j \wp(\xi)^j \right)
+ \left( \sum_{j=0}^{[(-p-3)/2]} b_j \wp(\xi)^j \right) \wp'(\xi),
\label{eqclasswp}
\end{eqnarray}
Since $\wp$ admits only one family, such a solution may exist
for any ODE.
It will be quite useful to take advantage of the value of the
singular part operator of $\wp(\xi)$,
\begin{eqnarray}
& &
\mathcal{D}= - \frac{\D^2}{\D \xi^2}.
\label{eqwpD}
\end{eqnarray}

For KS, the assumption to seek for solutions in the above class
\index{Kuramoto--Sivashinsky (KS)!equation}
\begin{eqnarray}
& & u=c_0 \wp' + c_1 \wp + c_2,\ c_0 \not=0.
\end{eqnarray}
together with the knowledge of the singular part operators 
(\ref{eqKSD}) and (\ref{eqwpD}),
first yields the correct values of $c_0$ and $c_1$,
\begin{eqnarray}
& & u= - 60 \nu \wp' - 15 b \wp + c_2,\ 
\end{eqnarray}
a truncation which defines the four determining equations
\begin{eqnarray}
& &
\left\lbrace
\begin{array}{ll}
\displaystyle{
b^2-16 \mu \nu=0,
}
\\
\displaystyle{
b \mu + 4 \nu c_2=0,
}
\\
\displaystyle{
b c_2 - 720 \nu^2 g_2=0,
}
\\
\displaystyle{
A+\frac{1}{2} c_2^2+\frac{15}{2} b^2 g_2 + 30 \mu \nu g_2 - 1080 \nu^2 g_3=0.
}
\end{array}
\right.
\end{eqnarray}
Their unique solution is (\ref{eqKSElliptic}).

\index{complex Ginzburg--Landau!(CGL3) equation}

For CGL3, the assumption that $M$ be in this class
\begin{eqnarray}
& & M=a_2 \wp + c_2,\ a_2 \not=0,
\end{eqnarray}
generates $10$ determining equations,
in the parameters and unknowns
$(a_2,c_2,g_2,g_3;d_r,d_i,s_r,c s_i,g_r,g_i)$.
The equation with the highest singularity degree
\begin{eqnarray}
& & ((d_i a_2)^2 - 9 a_2 d_r -18) (2 + a_2 d_r)=0
\end{eqnarray}
in which the vanishing of the second factor is forbidden,
is first solved as a linear equation for $d_r$
\begin{eqnarray}
& & d_r=\frac{(a_2 d_i)^2 - 18}{9 a_2}.
\end{eqnarray}
The next equation yields $c s_i=0$.
The next equations are successively solved for $g_r,g_2,g_3$,
and the elliptic discriminant $g_2^3-27 g_3^2$ is then divisible by the
unique remaining determining equation.
Therefore, one finds as unique solution the pulse
(\ref{eqCGL3Pulse}).

Finally, let us mention the Ansatz made for CGL5 \cite{Moores,AA1995}
\begin{eqnarray}
& &
A=a(x) e^{\displaystyle{i [- 2 \alpha \Log a(x) - \omega t]}},\
(\omega,\alpha,a) \in \mathcal{R},
\end{eqnarray}
which sets an \textit{a priori} constraint
between the amplitude and the phase
(similar to that made for CGL3 in Ref.~\cite{CT1989}),
together with the assumption that $a^2$ obeys a first order second degree
elliptic equation.
This allows one to retrieve (\ref{eqCGL5source}) in the particular case
$r_0=0,c=0$.

% ========================================================================
\section{Second class of methods: first order subequation}
\label{sectionSecondClass}

The requirement that the solution (\ref{eqFormalSolution})
be shared by the $N$th order ODE (\ref{eqODEReduced})
and the first order ODE (\ref{eqOrder1Autonomous})
characterizes, as recalled in the Appendix,
the singlevalued expressions $f$ as being elliptic
or degenerate elliptic (i.e.~trigonometric or rational), i.e.~the class
\begin{eqnarray}
& &
u=R(\wp',\wp) 
\longrightarrow
R(e^{k \xi})
\longrightarrow
R(\xi),
\end{eqnarray}
in which $R$ denotes rational functions and $\longrightarrow$ denotes the
degeneracy.
This class contains all the classes considered in previous sections
(polynomials in $\tanh$, in $(\tanh,\sech)$, in $(\sigma,\tau)$,
in $(\wp,\wp')$),
but it also contains in addition expressions like
(\ref{eqrationaltanh}).

The algorithm to obtain all the elliptic solutions combines
two pieces of information:

\begin{enumerate}
\item
a local one, a Laurent series representing the largest analytic solution
of the $N$-th order ODE near a movable pole-like singularity,

\item
a global one, the necessary form (\ref{eqsubeqODEOrderOnePP}) of 
(\ref{eqOrder1Autonomous}).

\end{enumerate}
by requiring that the Laurent series satisfies the first order subequation
(\ref{eqsubeqODEOrderOnePP}).

This provides the explicit form of the first order subequation $F(u,u')=0$.
Then one computes the solution $u=f(\xi-\xi_0)$ from this equation $F=0$.

The successive steps are \cite[Section 5]{MC2003}.

\begin{enumerate}

\item
Choose a positive integer $m$ and define the Briot and Bouquet
first order ODE
%(\ref{eqsubeqODEOrderOnePPwithp}),
\begin{eqnarray}
& &
F(u,u') \equiv
 \sum_{k=0}^{m} \sum_{j=0}^{[(m-k)(p-1)/p]} a_{j,k} u^j {u'}^k=0,\
a_{0,m}\not=0,
\label{eqsubeqODEOrderOnePPwithp}
\end{eqnarray}
in which $[z]$ denotes the integer part function.
The upper bound on $j$ implements the condition
$m(p-1) \le j p + k (p-1)$
that no term can be more singular than ${u'}^m$,
identically satisfied if $p=-1$.
The polynomial $F$ contains at most $(m+1)^2$ unknown constants $a_{j,k}$.

\item
Compute $\jmax$ terms of the Laurent series,
   with $\jmax$ slightly greater than the number of unknown constants
$a_{j,k}$.
\begin{eqnarray}
& &
u=\chi^p
 \left(\sum_{j=0}^{\jmax} M_j \chi^j+{\mathcal O}(\chi^{\jmax+1})\right),
\chi=\xi-\xi_0,
\label{eqLaurent}
\end{eqnarray}
where $p$ is
$-3$ for the KS equation (\ref{eqKSODE}),
$-2$ for the variable $\mod{A}^2$ of CGL3, etc.

\item
Require the Laurent series to satisfy the Briot and Bouquet ODE,
i.e.~require the identical vanishing of the Laurent series for the \LHS\
$F(U,U')$ up to the order $\jmax$
\begin{eqnarray}
& &
F \equiv \chi^{D} \left(\sum_{j=0}^{\jmax} F_j \chi^j
 + {\mathcal O}(\chi^{\jmax+1})
\right),\
D=m(p-1),\
\label{eqLaurentF}
\\
& &
\forall j\ : \ F_j=0.
\label{eqLinearSystemFj}
\end{eqnarray}
If it has no solution for $a_{j,k}$, increase $m$ and return to first step.

\item
For every solution,
integrate the first order autonomous ODE (\ref{eqsubeqODEOrderOnePPwithp}).

\end{enumerate}

Let us give two examples.

% ==========================================================================
\subsection{First order autonomous subequations of KS} % Primer, poubelle
\label{sectionKSsubeqFirstOrder}

\index{Kuramoto--Sivashinsky (KS)!equation}

The Laurent series of (\ref{eqKSODE}) is (\ref{eqKSODELaurent}).

In the second step, the smallest integer $m$ which allows a movable triple 
pole ($p=-3$) in (\ref{eqsubeqODEOrderOnePPwithp}) is $m=3$.
With the normalization $a_{0,3}=1$, the subequation contains ten coefficients,
which are first determined by the Cramer system of ten equations
$F_j=0,j=0:6,8,9,12$.
The first few are
\begin{eqnarray}
   F_0 & \equiv & -9 a_{0,3} + 40 \nu a_{4,0}=0,
\\ F_1 & \equiv & 9 b a_{0,3} + 12 \nu a_{1,2} -80 b \nu a_{4,0} =0,
\\ F_2 & \equiv &
(2120 b^2 + 2560 \mu \nu) \nu a_{4,0}
-(105 b^2 + 144 \mu \nu) a_{0,3}
-532 b \nu a_{1,2} 
-608 \nu^2  a_{2,1}
 =0,
\\ F_3 & \equiv &
 (5 b^2 + 72 \mu \nu) b a_{0,3} 
+ (137 b^2 + 240 \mu \nu) \nu a_{1,2} 
- (442 b^2 + 2656 \mu \nu) b \nu a_{4,0} 
\nonumber \\ & &
+ 608 b \nu^2 a_{2,1} 
+ 608 \nu^3 a_{3,0} 
=0.
\end{eqnarray}
The remaining infinitely overdetermined nonlinear system for $(\nu,b,\mu,A)$
contains as greatest common divisor (gcd) $b^2-16 \mu \nu$
(see Eq.~(\ref{eqKSConstraint16})),
which defines a first solution
\begin{eqnarray}
& &
\frac{b^2}{\mu \nu}=16,\
\left(u' + \frac{b}{2 \nu} u_s\right)^2
\left(u' - \frac{b}{4 \nu} u_s\right)
+\frac{9}{40 \nu}
\left(u_s^2 + \frac{15 b^6}{1024 \nu^4} + \frac{10 A}{3}\right)^2=0,\
\nonumber
\\
& &
u_s=u+\frac{3 b^3}{32 \nu^2}\cdot
\label{eqKSsubeqgenus1}
\end{eqnarray}
After division by this gcd,
the remaining system for $(\nu,b,\mu,A)$
admits exactly four solutions
(stopping the series at $j=16$ is enough to obtain the result),
namely the first three lines of Table \ref{TableKS},
each solution defining the same kind of subequation,
\begin{eqnarray}
& &
{\hskip -10.0 truemm}
b=0,\
%\frac{\nu k^2}{\mu}=\frac{11}{19},\
\nonumber
\\
& &
{\hskip -10.0 truemm}
\left(u' + \frac{180 \mu^2}{19^2 \nu}\right)^2
\left(u' - \frac{360 \mu^2}{19^2 \nu}\right)
+\frac{9}{40 \nu}
\left(u^2 + \frac{30 \mu}{19} u' - \frac{30^2 \mu^3}{19^2 \nu}\right)^2=0,\
\label{eqKSsubeqgenus0first}
\\
& &
{\hskip -10.0 truemm}
b=0,\
%\frac{\nu k^2}{\mu}=- \frac{1}{19},\
{u'}^3
+\frac{9}{40 \nu}
\left(u^2 + \frac{30 \mu}{19} u' + \frac{30^2 \mu^3}{19^3 \nu}\right)^2=0,\
\\
& &
{\hskip -10.0 truemm}
\frac{b^2}{\mu \nu}=\frac{144}{47},\
u_s=u-\frac{5 b^3}{144 \nu^2},\
\left(u' + \frac{b}{4 \nu} u_s\right)^3+\frac{9}{40 \nu} u_s^4=0,\
\\
& &
{\hskip -10.0 truemm}
\frac{b^2}{\mu \nu}=\frac{256}{73},\
u_s=u-\frac{45 b^3}{2048 \nu^2},\
\nonumber
\\
& &
{\hskip -10.0 truemm}
\left(u' + \frac{b}{8 \nu} u_s\right)^2
\left(u' + \frac{b}{2 \nu} u_s\right)
+\frac{9}{40 \nu}
\left(u_s^2+\frac{5 b^3}{1024 \nu^2}u_s + \frac{5 b^2}{128 \nu}u'\right)^2=0,\
\nonumber
\\
& &
{\hskip -10.0 truemm}
\label{eqKSsubeqgenus0last}
\end{eqnarray}

In order to integrate the two sets of subequations
(\ref{eqKSsubeqgenus1}),
(\ref{eqKSsubeqgenus0first})--(\ref{eqKSsubeqgenus0last}),
one must first compute their genus
\footnote{For instance with the Maple command \textit{genus} of the
package \textit{algcurves} \cite{MapleAlgcurves},
which implements an algorithm of Poincar\'e.},
which is one for (\ref{eqKSsubeqgenus1}),
    and zero for (\ref{eqKSsubeqgenus0first})--(\ref{eqKSsubeqgenus0last}).
Therefore (\ref{eqKSsubeqgenus1}) has an elliptic general solution,
listed above as (\ref{eqKSElliptic}),
and initially found \cite{FournierSpiegelThual,KudryashovElliptic}
by other methods.

As to the general solution of the four others
(\ref{eqKSsubeqgenus0first})--(\ref{eqKSsubeqgenus0last}),
this is the third degree polynomial (\ref{eqKSTrigo})
in $\tanh \frac{k}{2} (\xi-\xi_0)$.

These four solutions,
obtained for the minimal choice of the subequation degree $m$,
constitute all the analytic results currently known on (\ref{eqKSODE}).
For $m=4$, no additional solution is obtained \cite{YCM2003}.
The computation for $m=5$ is in progress.

% ==========================================================================
\subsection{First order autonomous subequations of CGL3} 
\label{sectionCGL3subeqFirstOrder}

\index{complex Ginzburg--Landau!(CGL3) equation}

We consider the variable $M=\mod{A}^2$, i.e.~$p=-2$.
The smallest value of $m$ is then $2$.
With the numerical values (\ref{eqCGL3Num}), 
the two Laurent series are
\begin{eqnarray}
& &
M_- = \chi^{-2} \left(
 -2+\frac{c}{5} \chi
 +\left(\frac{g_r}{3} -\frac{g_i}{6} -\frac{c^2}{200}\right) \chi^2
 + {\mathcal O}(\chi^3)\right),
\label{eqLaurentCGL3-}
\\
& &
M_+ = \chi^{-2} \left(
 4-\frac{2c}{5} \chi
 +\left(\frac{16g_r}{39} +\frac{4 g_i}{39} +\frac{19 c^2}{1300}\right) \chi^2
 + {\mathcal O}(\chi^3)\right).
\label{eqLaurentCGL3+}
\end{eqnarray}

The existence of two Laurent series, rather than just one,
is a feature which the subequation must also possess,
and this has the effect of setting the lower bound to $m=4$ instead of $2$.
Indeed,
the lowest degree subequations
\begin{eqnarray}
F_2 & \equiv & {M'}^2
 + {M'} (a_{1,1} M   + a_{0,1})
       + a_{3,0} M^3 + a_{2,0} M^2 + a_{1,0} M + a_{0,0}=0,
\label{eqsubeqODEOrderOnep2m2}
\\
F_3 & \equiv & {M'}^3
+ {M'}^2 (a_{1,2} M   + a_{0,2})
+ {M'}   (a_{3,1} M^3 + a_{2,1} M^2 + a_{1,1} M + a_{0,1})
\nonumber
\\
& &
+         a_{4,0} M^4 + a_{3,0} M^3 + a_{2,0} M^2 + a_{1,0} M + a_{0,0}
=0,
\label{eqsubeqODEOrderOnep2m3}
\end{eqnarray}
have the respective dominant terms
${M'}^2 + a_{3,0} M^3$ and ${M'}^3 + a_{3,1} {M'} M^3$,
which define only one family of movable double poles.

Let us nevertheless start with $m=2$,
for which (\ref{eqsubeqODEOrderOnep2m2}) can only be satisfied
by one series, e.g.~(\ref{eqLaurentCGL3-}),
thus preventing the full desired result to be obtained.
The six coefficients $a_{j,k}$ of (\ref{eqsubeqODEOrderOnep2m2})
are first computed as the unique solution of the linear system of
six equations $F_j=0,j=0,1,2,3,4,6$.
Then the $\jmax+1-6$ remaining equations $F_j=0,j=5,7:\jmax$,
which only depend on the fixed parameters $(g_r,g_i,c)$,
have the greatest common divisor (gcd) $3 g_i + 2 g_r + 3 c^2/50$,
and this factor defines the first solution 
((\ref{eqsubeqCGL3Hole-}) below).
After division par this gcd,
the system of three equations $F_j=0,j=5,7,8$,
provides two and only two other solutions,
see (\ref{eqsubeqCGL3Pulse-}) and (\ref{eqsubeqCGL3Front-}) below,
with the respective constraints $(c=0,g_i=0)$ and $(g_r=0,50 g_i-c^2=0)$,
and all the remaining equations $F_j=0,j \ge 9$, are identically satisfied.

Therefore, with this lower bound $m=2$,
one already recovers all the known first order subequations.
These are, with the series (\ref{eqLaurentCGL3-}),
\begin{eqnarray}
& &
\left(M' + \frac{c}{5} M + \frac{c^3}{250}\right)^2
+ 2 \left(M + \frac{c^2}{50}\right)
    \left(M - \frac{c^2}{50} - \frac{2}{3} g_r\right)^2 =0,\
3 g_i + 2 g_r + \frac{3 c^2}{50}=0,\
\label{eqsubeqCGL3Hole-}
\\
& &
{M'}^2 + 2 (M-g_r) M^2=0,\
c=0,\ g_i=0,\
\label{eqsubeqCGL3Pulse-}
\\
& &
\left(M' + \frac{c}{5} M\right)^2 + 2 M^3=0,\
g_r=0,\ g_i - \frac{c^2}{50}=0.
\label{eqsubeqCGL3Front-}
\end{eqnarray}

Finally, for each of the three subequations,
the fourth step finds a zero value for the genus
and returns the general solution as a rational function of
$e^{a (\xi-\xi_0)}$,
which basic trigonometric identities then allow to convert to the
second degree polynomials in $(k/2) \tanh k (\xi-\xi_0)/2$
listed in (\ref{eqCGL3Hole-})--(\ref{eqCGL3Front-}).

Similarly, with the other series (\ref{eqLaurentCGL3+}), one obtains
\begin{eqnarray}
& &
\left(M' + \frac{c}{5} M - \frac{c^3}{500}\right)^2
-   \left(M - \frac{c^2}{100}\right)
    \left(M + \frac{c^2}{100} - \frac{2}{3} g_r\right)^2 =0,\
3 g_i - g_r + \frac{3 c^2}{80}=0,\
\label{eqsubeqCGL3Hole+}
\\
& &
{M'}^2 - M \left(M -\frac{2}{3} g_r\right)^2=0,\
c=0,\ 3 g_i - g_r =0,\
\\
& &
\left(M' + \frac{c}{5} M\right)^2 - M^3=0,\
g_r=0,\ g_i - \frac{c^2}{50}=0.
\label{eqsubeqCGL3Front+}
\end{eqnarray}

With the correct two-family lower bound $m=4$,
which corresponds to $18$ unknowns $a_{j,k}$ and
at least $24$ terms in the series,
we have checked that there is no solution other than the above three.
This situation is quite similar to the absence of solution in the class
(\ref{eqCGL3TwoFamiliesSolution}),
and it just reflects the difficulty of the CGL3 equation.

The case $m=8$
($60$ unknowns $a_{j,k}$ and at least $66$ terms in the series)
is currently under investigation
but preliminary results seem to indicate the absence of any new solution,
and we are now automating the computer algebra program
in order to handle much larger values of $m$.

% ==========================================================================
\subsection{Domain of applicability of the method} 
\label{sectionDiscussion}

As we have seen,
the subequation method contains the truncation methods 
and its cost is minimal since the main step is a linear computation.

The two key assumptions behind this ``subequation method'' are,
\begin{enumerate}

\item
a Laurent series should exist,

\item
a first order autonomous algebraic subequation should exist.

\end{enumerate}

Its best applicability is therefore nonintegrable
$N$-th order autonomous nonlinear ODEs
admitting a Laurent series which only depends on one movable constant,
such as the CGL3 ODE (\ref{eqCGL3Order3})
or the traveling wave reduction 
(\ref{eqKSODE}) of the Kuramoto-Sivashinsky equation
\cite{CM1989,YCM2003}.

Two examples of inapplicability are
\begin{enumerate}

\item
the Lorenz model,
in which the Laurent series generically does not exist
and has to be replaced by a psi-series \cite{Segur},

\item
the autonomous ODE $u'''-12 u u' -1=0$,
which admits the first Painlev\'e transcendent as its general solution,
a case in which no first order subequation exists.

\end{enumerate}

% ========================================================================
\section{Conclusion}

How do these two classes of methods (truncations, first order subequations)
really compare,
independently of the amount of computation involved?

Let us first recall a preliminary, classical result.

The class (\ref{eqTruncationuOneFamily}) of polynomials of
degree $-p$ in $\tanh$ obeys a first order equation of degree $m=-p$.
For instance, given the polynomial
\begin{eqnarray}
& &
u=\tanh^2+2 a\tanh+b,
\end{eqnarray}
this amounts to eliminate $\tanh$ between the two algebraic equations
\begin{eqnarray}
& &
\left\lbrace
\begin{array}{ll}
\tanh^2+2 a\tanh+b-u=0,
\\
2(\tanh+ a)(1-\tanh^2)-u'=0,
\end{array}
\right.
\label{eqtanh2elimin}
\end{eqnarray}
which results in
\footnote{
This formula, due to Sylvester,
expresses the resultant of two polynomials of degrees $m$ and $n$
as a determinant of order $m+n$.
}
\begin{eqnarray}
& &
\left|\matrix{
 0 &  0   &  1   & 2 a    &   b-u  \cr
 0 &  1   &  2 a &   b-u  & 0      \cr
 1 &  2 a &  b-u & 0      & 0      \cr
 0 & -2   & -2 a & 2      & 2 a-u' \cr
-2 & -2 a &  2   & 2 a-u' & 0      \cr
 }
\right|
 \\
 & &
 =
(u'-4 a (u-b+a^2))^2 -4 (u-b+ 2 a^2-1)^2 (u-b+a^2)=0,
\end{eqnarray}
an equation with degree $m=2=-p$, having genus zero.

Similarly, the class of polynomials of global degree $-p$ in $(\tanh,\sech)$
or $(\sigma,\tau)$ obeys a first order ODE with degree $m=-2p$.

Last, the class (\ref{eqclasswp}) of polynomials of $(\wp,\wp')$ 
of singularity degree at most equal to $p$
obeys a first order ODE with degree $m=-p$.

Therefore,
given a value of $p$ (the singularity degree of the ODE)
and a truncation considered in section
\ref{sectionFirstClass},
there exists a value of $m$ (either $-p$ or $-2p$)
at which the result of the truncation can be found by the method of
first order subequations.

Conversely, given a value of $m$ (the degree of a first order subequation),
the class of solutions of the method of first order subequations
is made of the rational functions
(of $(\wp,\wp')$ or of $e^{k \xi}$ i.e. $(k/2)\tanh k \xi/2$),
a class richer than the polynomials.

This proves the \textit{identity} of the two classes of methods,
provided the truncations assume rational functions instead of
polynomials.

However, from the practical point of view of the amount of
computation involved,
the increasing order of difficulty seems to be

\begin{enumerate}

\item
Truncations of polynomials.

\item
First order subequations.

\item
Truncations of rational functions.

\end{enumerate}

How does this compare with the approach of Chow (see e.g. \cite{Chow2002})
to find solutions of PDEs in terms of elliptic functions?
To be definite, let us consider a PDE in $(x,t)$.
The solutions which do depend on both $x$ and $t$
(i.e. which do not satisfy some ODE)
are richer than those here described.
As to the solutions of the solitary wave type $f(x-ct)$,
the method of Chow belongs to the first class of methods,
i.e.~it may or it may not find the most general elliptic solution 
which exists.

% ==========================================================================
\section*{Acknowledgments}

We gratefully acknowledge the Tournesol grant T2003.09.

\printindex

% ========================================================================
\section{Appendix. Classical results on first order autonomous equations}
\label{sectionClassicalResultsFirstOrderAutonomous}

The following results were
mainly obtained by Briot and Bouquet, Fuchs, Poincar\'e
and put in final form by Painlev\'e \cite[pages 58--59]{PaiLecons}.

\textbf{Theorem}.
Given an algebraic first order autonomous ODE (\ref{eqOrder1Autonomous}),
whose general solution is therefore (\ref{eqFormalSolution}),
the following properties are equivalent.

\begin{enumerate}

\item
Its general solution is singlevalued.

\item
Its general solution is an elliptic function, possibly degenerate.

\item
The genus of the algebraic curve (\ref{eqOrder1Autonomous}) is one or zero.

\item
There is equivalence between the knowledge of $f$ and that of $F$,

\item
There exist a positive integer $m$ and $(m+1)^2$
complex constants $a_{j,k}$, with $a_{0,m}\not=0$, such that
the polynomial $F$ of two variables has the necessary form
\begin{eqnarray}
& &
F(u,u') \equiv
 \sum_{k=0}^{m} \sum_{j=0}^{2m-2k} a_{j,k} u^j {u'}^k=0,\ a_{0,m}\not=0.
\label{eqsubeqODEOrderOnePP}
\end{eqnarray}

\item
If the genus is one,
there exist two rational functions $R_1,R_2$,
such that the general solution is
\begin{eqnarray}
& &
u=R_1(\wp) + \wp' R_2(\wp),
\label{equWeierstrassForm}
\end{eqnarray}
in which $\wp=\wp(\xi - \xi_0,g_2,g_3)$ is the Weierstrass elliptic function
characterized by (\ref{eqWeierstrass}).

If the genus is zero,
there exists a (possibly zero) constant $a$ and a rational function $R$
such that the general solution is
\begin{eqnarray}
& &
u=R(e^{a \xi}),
\end{eqnarray}
with the degeneracy $u=R(\xi)$ in case $a$ is zero.

\end{enumerate}

% ***************************************************************** References

\vfill \eject
\end{document}